\documentclass[prb, twocolumn, superscriptaddress, showpacs,floatfix]{revtex4-1}  
\usepackage{hyperref}
\usepackage{amssymb}
\usepackage{amsmath}
\usepackage{float}
\usepackage{bbm}
\usepackage{graphicx}
\usepackage{epsfig}
\usepackage{epstopdf}
\usepackage[usenames]{color}

\begin{document}
\bibliographystyle{apsrev4-1}

\title{Energetics of superconductivity in the two dimensional Hubbard model}

\author{E. Gull}
\affiliation{Department of Physics, University of Michigan, Ann Arbor, Michigan 48109, USA}

\author{A. J. Millis }
\affiliation{Department of Physics, Columbia University, New York, New
  York 10027, USA}

\date{\today }

\begin{abstract}
The energetics of the  interplay between superconductivity and the pseudogap in high temperature superconductivity is examined using the eight-site dynamical cluster approximation to the two dimensional Hubbard model. Two regimes of superconductivity are found: a weak coupling/large doping regime in which the onset of  superconductivity causes  a reduction in potential energy and an increase in kinetic energy, and a strong coupling regime in which superconductivity is associated with an increase in potential energy and decrease in kinetic energy. The crossover between the two regimes is found to coincide with the boundary of the normal state pseudogap, providing further evidence of the unconventional nature of superconductivity in the pseudogap regime. However the absence, in the strongly correlated but non-superconducting state,  of discernibly nonlinear response to an applied pairing field, suggests that resonating valence bond physics is not the origin of the kinetic-energy driven superconductivity. 
\end{abstract}

\pacs{
74.20.-z,%Theories and models of superconducting state
71.10.Fd,%Lattice fermion models (Hubbard model, etc.)
74.25.Dw,%Superconductivity phase diagrams
74.72.−h,%Cuprate superconductors
%
%These could also apply
%71.27.+a,%Strongly correlated electron systems; heavy fermions 
%71.10.Hf,%Non-Fermi-liquid ground states, electron phase diagrams and phase transitions in model systems
%74.72.Kf,%Cuprates/Pseudogap regime 
%71.30.+h,%Metal-insulator transitions and other electronic transitions
%71.10.−w,%Theories and models of many-electron systems
%74.20.mn,%Nonconventional mechanisms 
}

\maketitle

The high transition temperature superconductivity exhibited by layered copper-oxide materials has been an important topic in condensed matter physics since its discovery in 1986.\cite{Bednorz86} Broadly speaking, two views  are currently held. One is that despite the various anomalous features of the materials the superconductivity may be understood in more or less conventional Bardeen-Cooper-Schrieffer (BCS) terms as arising from the exchange of a pairing (`glue') particle, most likely of magnetic origin.\cite{Monthoux92} An alternative view  is that the superconductivity is an intrinsic property of a strongly correlated state of matter that should not be interpreted as arising from the exchange of a well-defined excitation.\cite{Anderson87}  

The issue may be cast in energetic terms. In the conventional BCS view, the driving force for superconductivity is in essence a reduction of potential energy: by forming the superconducting state the electrons can take greater advantage of an attractive term in an interparticle potential. Changing the wave function to reduce the potential energy however costs kinetic energy, so that  in the weak coupling limit the change from normal to superconducting states leads to an increase in the kinetic energy.\cite{Chester56} In an alternative view,\cite{Anderson87} the driving force for superconductivity is an optimization of kinetic energy: by forming the superconducting state the electrons can move more easily through the crystal despite their need to avoid the other electrons. In this case  going from the normal to the superconducting state lowers  the kinetic energy  and one expects that the potential energy increases. 

The repulsive-U Hubbard model on the two dimensional square lattice is widely believed \cite{Anderson87,Scalapino07,Lee07} to contain the essential physics of high-$T_c$ copper-oxide superconductivity. It  is defined by  the Hamiltonian
%(written here in a mixed momentum/position representation)
\begin{equation}
H=\sum_{k\sigma}(\varepsilon_k-\mu)c^\dagger_{k\sigma}c_{k\sigma}+U\sum_in_{i\uparrow}n_{i\downarrow}
\label{H}
\end{equation}
Here $i$ labels the sites in a lattice and $k$ a momentum in the corresponding Brillouin zone. 
%Each site is assumed to contain a single orbital which may hold up to one electron of spin $\sigma=\uparrow$   and one electron of spin $\sigma=\downarrow$.  The interaction $U$ favors ($U<0$) or disfavors ($U>0$)  configurations with two electrons on a site.  
The two dimensional repulsive ($U>0$) version of the  model has been shown rigorously to have a d$_{x^2-y^2}$  superconducting ground state in at least some regions of the $U,n$ phase diagram.\cite{Zanchi96,Raghu10,Maier05_dwave}

\begin{figure}[t]
\begin{center}
\includegraphics[angle=0.0, width=0.85\columnwidth]{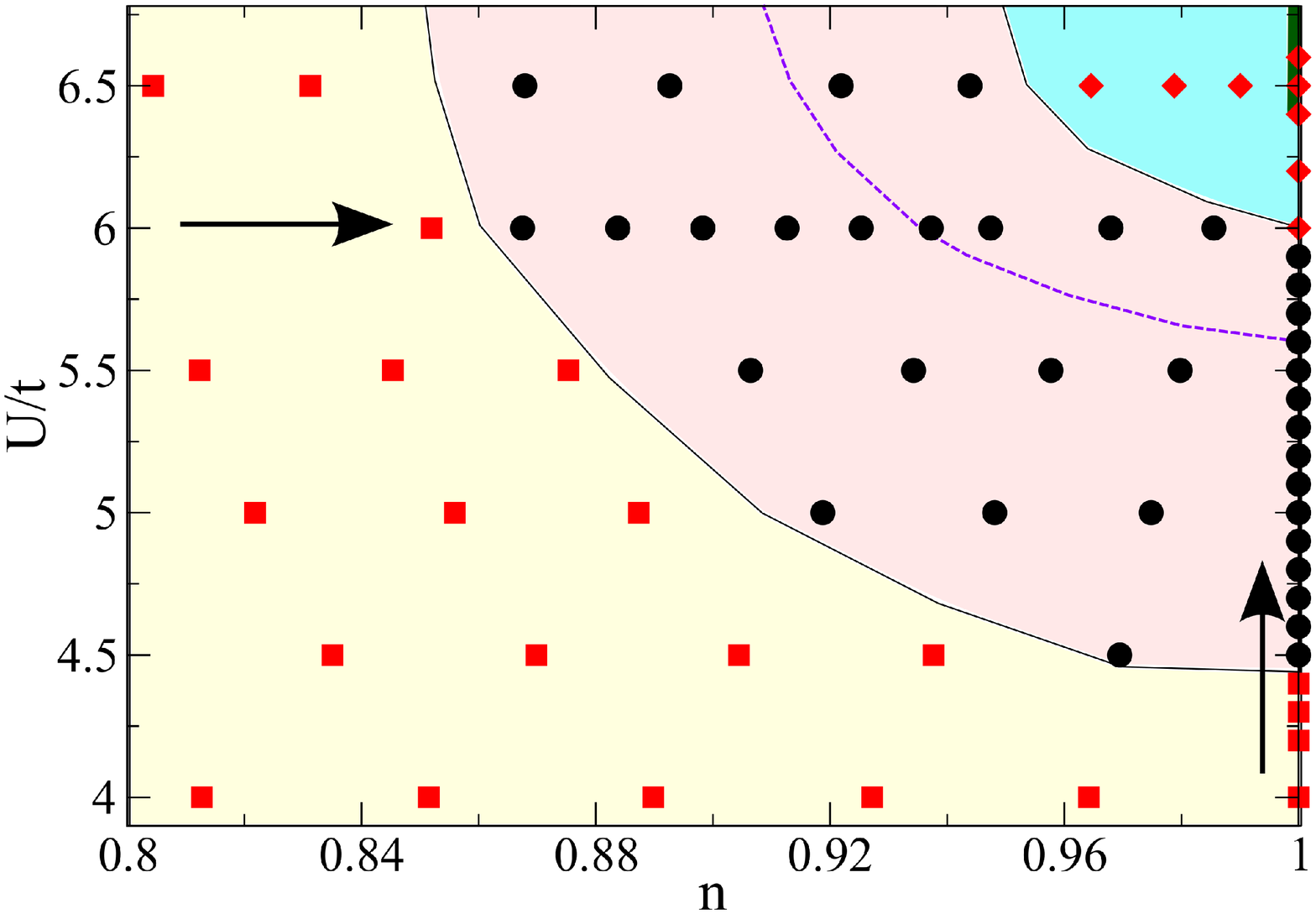}
\caption{(Color online) Phase diagram of two dimensional square lattice Hubbard model in plane of density $n$ and interaction strength $U/t$ at inverse temperature $\beta=60/t$ as obtained in 8-site cluster dynamical mean field theory. Mott insulator at half filling for $U/t \gtrapprox 6.4$ indicated by heavy bar (green online); superconducting region indicated by circles (black online), pseudogapped but non-superconducting region, diamonds (blue online) and Fermi liquid non-superconducting state by squares (yellow online). Boundary of normal state pseudogap, defined as in Ref.~\onlinecite{Gull09}, indicated as dashed line (purple online). Trajectories along which the energy is computed are shown as arrows. 
\label{phasediagram}}
\end{center}
\end{figure}

In this paper we investigate the electronic energy $E=\langle H\rangle$, 
%defined as the expectation value of the Hamiltonian  and 
%\begin{equation}E=\left<H\right>\label{Edef}\end{equation} It may be 
decomposed into kinetic $K$ and potential $V$ terms
%and particle number $N$ terms 
as $E=K+V$ with
%\begin{eqnarray}
\begin{eqnarray}
K\!\!&=&\!\!\sum_{k\sigma}\varepsilon_k\langle c^\dagger_{k\sigma}c_{k\sigma}\rangle=2T\sum_{k,n}(\varepsilon_k-\mu)\text{Tr}\left[\tau_3G(k,\omega_n)\right]
\label{Kdef} \\
V\!\!&=&\!\!U\sum_i\left<n_{i\uparrow}n_{i\downarrow}\right>=2T\sum_{k,n}\text{Tr}\left[\Sigma(k,\omega_n)G(k,\omega_n)\right]
\label{Udef}
%\\
%N&=&\mu\sum_{k\sigma}\left<c^\dagger_{k\sigma}c_{k\sigma}\right>=2T\mu\sum_{k,n}Tr\left[G(k,\omega_n)\right]
%\label{Ndef}
\end{eqnarray}
%\end{eqnarray}
In the second equality we have used standard formulae to 
reexpress the expectation values in terms of the Nambu matrix Matsubara frequency electron Green function G and self energy $\Sigma$.
%reexpress the expectation values in terms of the  Matsubara frequency electron Green function $G$ and self energy $\Sigma$. We  %assume for definiteness a paramagnetic state but 
%write the formulae in Nambu matrix notation to allow for the possibility of superconductivity. 

The energetics of superconductivity have been previously studied. One important class of approaches has used variational wavefunctions, often starting from a single Slater determinant  with doubly occupied sites then being projected out.\cite{Gros88,Yokoyama88,Becca00,Paramekanti01,Paramekanti04,Yokoyama04,Ogata12a,Ogata12b} Information about pairing comes from comparing results obtained from free fermion and BCS-paired starting points. These works indicated that pairing was present for dopings from $x=0$ to $x\sim 0.25$ and that over most of the phase diagram the kinetic energy of the paired state was lower than that of the unpaired state.  However, variational results are constrained by the choice of variational space; in particular by the choice of projective BCS-type wave functions.

Another important class of theoretical approaches involves phenomenological spin-fermion models.\cite{Haslinger03a,Haslinger03b,Yanase05,Benfatto06,Norman07,Marsiglio08,Maiti10} In these approaches it is assumed that the important physics arises from the interaction of electrons with spin fluctuations (treated as bosons but with boson self-energy effects arising from coupling to fermions playing a crucial role). These models are amenable to semi-analytic treatment. Their analysis revealed that  in the strong coupling limit  the superconducting state could have lower kinetic energy than the normal state. However, these models do not fully capture the strong correlation effects associated with the Mott transition or the formation of the pseudogap, and  rely on assumptions about the most physically relevant interactions.

We  use the dynamical cluster approximation (DCA) version of dynamical mean field theory \cite{Maier05} to evaluate Eqs.~[\ref{Kdef},\ref{Udef}] for  the two dimensional repulsive-U Hubbard model with $\varepsilon_k=-2t(\cos k_x+\cos k_y)$. In the DCA the Brillouin zone is tiled with  $N$ patches and  the electron self energy is taken to be piecewise constant, with a different value in each sector of momentum space. The sector self energies are obtained from the solution of an auxiliary quantum impurity model with parameters fixed by the Hubbard interaction and a self-consistency condition discussed in detail in Ref.~\onlinecite{Maier05}. The method yields a d$_{x^2-y^2}$ superconducting state.\cite{Maier00,Lichtenstein00,Maier06,Scalapino07,Maier07,Maier07B,Civelli08,Kancharla08,Maier08,Civelli09,Civelli09b,Sordi12,Gull12} For the Hubbard model the method becomes exact as $N\rightarrow\infty$ and considerable evidence is now available \cite{Kozik10,Gull10_clustercompare,Fuchs11,Sakai12} concerning the status of the finite $N$ results achievable numerically. Here we study the case $N=8$, which has been shown to be   large enough for the results to be representative of the infinite cluster size limit \cite{Maier05_dwave} but small enough to enable calculations of the necessary accuracy.\cite{Gull10_clustercompare} 

%The energetics of superconductivity have been previously studied. Maier and co-workers presented calculations using  the $N=4$ DCA approximation with an approximate `impurity solver' based on the noncrossing approximation (NCA). They concluded that in the Hubbard model  the transition to the superconducting state always involved lowering the kinetic energy \cite{Maier04}. On the other hand, in a subsequent 4-site cluster study of the $t-J$ model, Haule and Kotliar concluded that while the dominant contribution to the energy change across the normal-superconducting transition was the change in the expectation value of the $J$ term, the kinetic energy variation dependence depended on doping: on the overdoped side the kinetic energy increased on entering the superconducting phase whereas on the underdoped side of the phase diagram the  kinetic energy decreased \cite{Haule07}.  Singh \cite{Singh07} has questioned the relevance of computations based on the t-J model, because of apparent violations of the virial theorem which may be traced back to the fact that some parts of the electron kinetic energy are included in the `J' coupling.  As we shall see our results differ from those reported in Refs.~\onlinecite{Maier04,Haule07}.

We obtained the superconducting kinetic and potential energies $KE_S$ and $PE_S$ from superconducting solutions obtained as described in Ref.~\onlinecite{Gull12} and the normal state energies $KE_N$ and $PE_N$ by solving the DMFT equations in the paramagnetic phase with the same code but  subject to the constraint that the anomalous ($\langle cc\rangle$) terms in the Green function and self energy vanished.  Our results are obtained using  the CT-AUX version \cite{Gull08}  of the continuous-time quantum Monte Carlo method \cite{Gull11} with submatrix updates \cite{Gull10_submatrix} and an extension to superconductivity.\cite{Gull12} The energy differences are found to be very small and careful attention to the high frequency behavior is required for reliable results. The submatrix  methods are essential in obtaining data of the requisite accuracy.  
%The results were cross-checked by an independent code that computed the normal-superconducting differences directly (the high frequency convergence issues are much less severe in this case). 

Fig.~\ref{phasediagram} shows the phase diagram obtained from the $N=8$ DCA method in the interaction strength and doping plane \cite{Gull12} along with two arrows indicating the parameter-space trajectories along which energies are computed in this paper. At  $U\gtrsim6.4t$ and carrier concentration $n=1$ per site the approximation yields a paramagnetic (`Mott') insulating state which is at lower temperatures unstable to antiferromagnetism. As electrons are removed the state evolves to a conventional Fermi liquid metal via an intermediate `strange metal' phase characterized by a `pseudogap', a suppression of electronic density of states in the $(0,\pi)$ region of the Brillouin zone.\cite{Huscroft01,Parcollet04,Macridin06,Gull09,Werner098site,Sakai09,Gull10_clustercompare,Sakai10,Sordi10,Yang11,Sordi11} Superconductivity is found in a strip,\cite{Gull12} near to the Mott insulator but separated from it by a region of pseudogapped but nonsuperconducting states.\cite{Yang11}  At carrier concentration $n=1$ (vertical arrow) the ground state of the model is believed to be antiferromagnetic at all $U$. The $n=1$ results  were obtained by suppressing long-ranged antiferromagnetic order (although short-ranged antiferromagnetic  correlations are still present) and are representative of the properties of a metastable state. They are included because the qualitative properties are seen to be the same as in the doping-driven transition but  the particle-hole symmetry at $n=1$ permits the acquisition of much higher quality data, enabling a clearer view of the phenomena. 

\begin{figure}[tb]
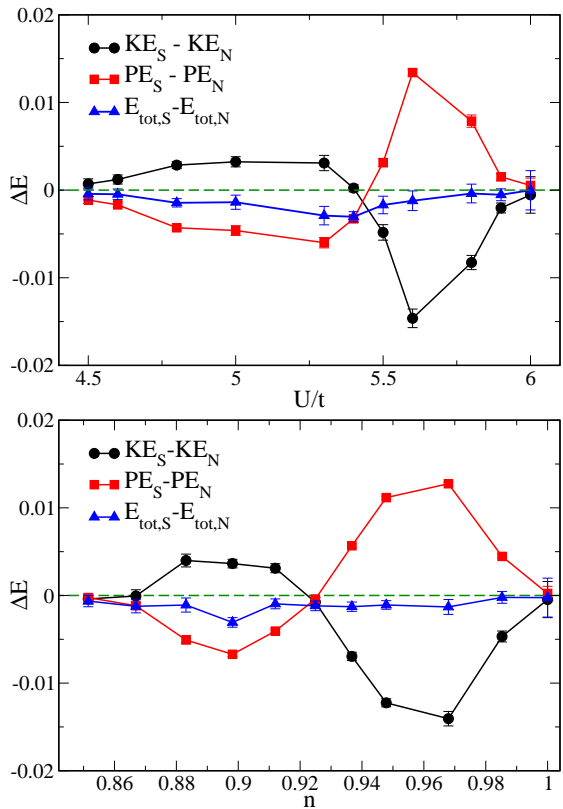

\begin{center}
\includegraphics[angle=0.0, width=0.85\columnwidth]{energyu}
\includegraphics[angle=0.0, width=0.85\columnwidth]{energyn}
\caption{(Color online) Differences in total, kinetic and potential energies (per site, in units of hopping $t$) between normal and superconducting states, obtained as described in the text at density $n=1$ varying interaction strength (upper panel) and as function of density at fixed interaction strength $U=6t$ (lower panel).
}
\label{energy}
\end{center}
\end{figure}

The two panels of Fig.~\ref{energy} show the energy differences obtained by subtracting the superconducting and normal state energies computed at inverse temperature $\beta=60/t$ along the two parameter-space trajectories shown by the arrows in Fig.~\ref{phasediagram}, i.e.  crossing the superconducting region by varying the interaction strength or varying the carrier concentration. The results obtained along the two trajectories are remarkably similar, although the absence of a fermion sign problem at $n=1$ means we are able to obtain much better statistics in this case. The condensation energy is of order $0.001t$, although the changes in kinetic and potential energy separately are typically much larger, especially in the pseudogap regime. %We see that the superconducting condensation energy is largest at the boundary of the pseudogap regime, and that at this point the energetics changes. 

For interactions or carrier concentrations smaller than required to produce a normal state pseudogap,\cite{Gull09,Werner098site,Gull10_clustercompare} the energetics are consistent with the standard expectations of weak coupling superconductivity: as the material enters the superconducting state the potential energy decreases and the kinetic energy increases. The boundary of the normal state pseudogap marks a significant change in the energetics of superconductivity: once the pseudogap regime is entered, the kinetic energy decreases and the potential energy increases on entering the superconducting state. Further, inside the pseudogap regime the superconducting/normal changes in potential and kinetic energy become much larger in magnitude, showing that the onset of superconductivity leads to a significant re-organization of the energetics of the pseudogap states. The change in character of the superconductivity at the pseudogap line is consistent with the finding of \textcite{Yang11} that the superconductivity exists in a dome with the maximal transition temperature occurring where the superconducting and pseudogap phase boundaries intersect.

%\begin{figure}[t]\begin{center}\includegraphics[angle=0.0, width=0.8\columnwidth]{Tc_and_Econd_vs_U}
%\includegraphics[angle=0.0, width=0.8\columnwidth]{Tc2vsEcond}
%\caption{(Color online) Square of transition temperature $T_c$ (multiplied by $15$) and condensation energy plotted against interaction strength at carrier concentration $n=1$.}\label{TcvsEcond}\end{center}\end{figure}

%In weak-coupling theory the condensation energy scales as the square of the transition temperature\cite{Chester56}.  The error bars in our condensation energy are relatively large, so only a qualitative consistency test is possible, but it appears that over the entire parameter range the condensation energy tracks the square of the transition temperature. 

Our results differ from previous dynamical mean field analyses. Ref.~\onlinecite{Maier04} ($N=4$ study of the Hubbard model) and Ref.~\onlinecite{Haule07} ($N=4$ study of the t-J model, with an additional `EDMFT' approximation) found that most of the energy gain on entering the superconducting state came from changes in the interaction term, although Ref.~\onlinecite{Haule07} found that the behavior of the kinetic energy was different at large than at small doping. Three possible origins for the discrepancy are the use of the non-crossing approximation  (``NCA'') impurity solver in Refs.~\onlinecite{Maier04,Haule07} rather than the numerically exact CT-QMC method,  the use of the $N=4$ approximation, rather than the  $N=8$ approximation used here, and the study of the t-J rather than Hubbard model in Ref.~\onlinecite{Haule07}. Singh \cite{Singh07} has questioned the relevance of computations based on the t-J model, because of apparent violations of the virial theorem which may be traced back to the fact that some parts of the electron kinetic energy are included in the `J' coupling.  

The `potential energy-driven' nature of the superconductivity  found at larger dopings and at weak couplings is consistent with the notion that in these regimes the superconductivity is relatively conventional. The change in energetics as the pseudogap boundary is crossed suggests that at stronger couplings or lower dopings the superconductivity becomes unconventional. One influential model of unconventional superconductivity is the resonating valence bond (RVB) idea of Anderson \cite{Anderson87} which was motivated in part by the possibility that the physics of the cuprates could be understood in terms of a very strong coupling limit of the Hubbard model. There,% In the strong coupling limit,  
configurations with two electrons on a site  could be projected out so that the only important term in the energy was the kinetic energy term and superconductivity (and indeed all other interesting physics) is necessarily `kinetic energy driven'. 

\begin{figure}[tb]
\begin{center}
\includegraphics[angle=0.0, width=0.85\columnwidth]{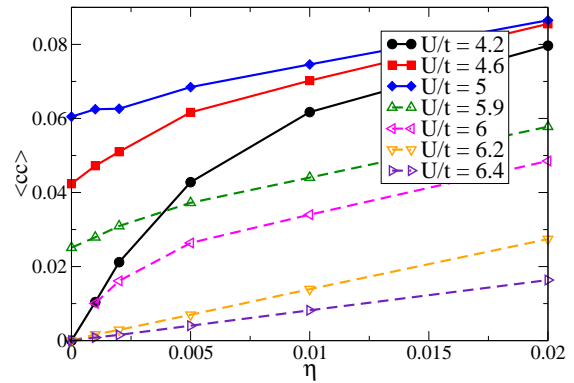}
\caption{(Color online) Anomalous expectation value in sector $K=(0,\pi)$ plotted against pairing field $\eta$ at doping $x=0$ for interaction strengths indicated.
}
\label{ccvseta}
\end{center}
\end{figure}
Anderson's original RVB idea, and subsequent recognition of an $SU(2)$ symmetry in the RVB wave function which might be weakly broken by doping or small finite $U$ corrections  \cite{Affleck88} implied that superconducting correlations were present (but not active) in the Mott insulating state and the strongly correlated but not superconducting state which separates the superconducting and insulating regimes in Fig.~\ref{phasediagram}.  To test this hypothesis we applied a pairing field $\eta_K(c^\dagger_{K\uparrow}c^\dagger_{K\downarrow}+c_{K\uparrow}c_{K\downarrow})$ in our calculations and computed the effect on the superconducting order parameter $\langle c_{K\uparrow}c_{K\downarrow}\rangle$. We expect that if a near-$SU(2)$ symmetry existed, then applying a small pairing field to a state which is  non-superconducting but is near the phase boundary would provide a rapid increase in the pairing amplitude, which would then saturate to a value characteristic of the superconducting state.  Fig.~\ref{ccvseta} shows that this is not the case. On the weak coupling side ($U=4.2$), applying a pairing field leads to the behavior expected near a second order phase transition: a rapid increase in $\langle cc\rangle$ reflecting the enhanced susceptibility, followed by a saturation to values similar to those found in the nearby superconducting state. However, on the strong coupling side the situation is different.  Just at the phase boundary $U=6.0$ the situation is similar to that found at weak coupling, but for any larger $U$ the $\langle cc\rangle$ vs $\eta$ curve is linear with small, weakly U-dependent slope. The similarity of the $U=6.2$ and $6.4$ results, and the difference of both of these to the $U=4.2$ trace, indicates that precursor effects are very weak as the superconducting phase is approached from the pseudogap indicating that the pseudogap state has no strong  tendency towards superconductivity. We infer from this calculation that the origin of the kinetic energy-driven behavior is not  a signature of pairing correlations pre-existing in the wave function. 

\begin{figure}[tb]
\begin{center}
\includegraphics[angle=0.0, width=0.85\columnwidth]{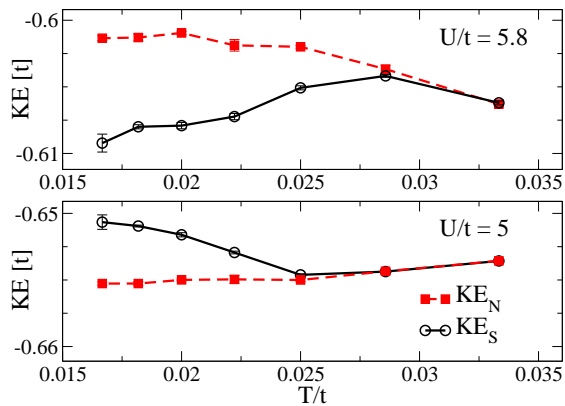}
\caption{(Color online) Temperature dependence of kinetic energy at $n=1$ for $U=5.0$ (lower panel) and $U=5.8$ (upper panel) in normal (filled squares, red dashed line) and superconducting state (open circles, black solid line).
}
\label{KeofT}
\end{center}
\end{figure}

It is interesting to consider the normal - superconducting energy differences in the context of the energetics of the pseudogap state itself. The two panels of Fig.~\ref{KeofT}  show the temperature dependence of the kinetic energy computed for a relatively weak coupling, $U=5.0t$, (lower panel) and  relatively strong coupling, $U=5.8t$, (upper panel). We see that in the weak coupling case, the kinetic energy decreases as the temperature is lowered, and the onset of superconductivity reverses this decrease, while in the strongly coupled case the kinetic energy increases as temperature is lowered but the onset of superconductivity again reverses the temperature dependence. 

We finally consider the observability of these effects. Norman and co-workers noted that the difference between normal-state and superconducting state photoemission spectra could be analyzed to obtain estimates of the normal-superconducting  change both in total and in kinetic energy \cite{Norman00}  although the analysis is complicated by the need to accurately monitor small changes occurring over wide energy and momentum ranges. Specific heat data can also be used to infer the condensation energy, although care must be taken both to extrapolate the normal state to temperatures less than the transition temperature and to include fluctuation effects.\cite{Marel02} In narrow-band systems such as cuprates, an approximate relation between the kinetic energy and the frequency integral of the  optical conductivity exists.\cite{Maldague77,Baeriswyl87,Millis90} Interestingly the idea of examining superconductivity-induced changes in the optical integral seems  to have entered the high-T$_c$ literature first in the context of the `hole superconductivity' model of Hirsch \cite{Hirsch93,Hirsch00} and then by Chakravarty and collaborators \cite{Chakravarty98}  in relation to the interlayer coherence mechanism of Anderson.\cite{Anderson98} %At present neither model has wide  acceptance as a  description of superconductivity in the copper-oxide compounds, but  
These works  motivated several experimental groups to examine changes in optical conductivity across the normal-superconducting phase boundary.\cite{Molegraaf02,Santander02,Santander04,Deutscher05,Carbone06} Unfortunately, the value of the optical spectral weight depends on the frequency up to which the conductivity is integrated, and the appropriate upper cutoff may be different in the normal and superconducting state;\cite{Norman07,Marsiglio08} also even if the conduction band contribution to the optical sum rule could be determined, the relation between this and the kinetic energy is only approximate, and the errors involved in the approximation may be different in the normal and superconducting state.\cite{Maiti10} As can be seen from Fig.~\ref{KeofT}, the temperature dependent changes are only on the $1\%$ level.

In summary, our results indicate that the nature of the superconductivity depends crucially on location in the phase diagram. For the high doping/weak correlation side of the superconducting region, the energetics of superconductivity appear essentially conventional: on transition to the superconducting state the potential energy decreases and the kinetic energy increases. On the low doping/strong correlation side of the superconducting region, the energetics appears unconventional: on entering the superconducting state the kinetic energy decreases and the potential energy increases. However, we do not find any indication that the nonsuperconducting  pseudogap state has any significant pairing correlations, casting doubt on an RVB interpretation of the pseudogap.  Interestingly, the crossover between the two regimes occurs essentially at the point at which the pseudogap becomes visible in normal state quantities and also interestingly the changes in the individual components of the energy become much larger in this unconventional regime, suggesting that superconductivity causes a substantial rearrangement of the pseudogap electronic state. In other words, the superconductivity and pseudogap are competing phases. 

{\it Acknowledgments:} AJM was supported by NSF-DMR-1006282. We thank A. Chubukov and M. Norman for helpful conversations. 
%This research used resources of the National Energy Research Scientific Computing Center, which is supported by the Office of Science of the U.S. Department of Energy under Contract No. DE-AC02-05CH11231. 
A portion of this research was conducted at the Center for Nanophase Materials Sciences at Oak Ridge National Laboratory and at the National Energy Research Scientific Computing Center (DE-AC02-05CH11231), which are supported by the Office of Science of the U.S. Department of Energy. Our continuous-time quantum Monte Carlo codes are based on ALPS.\cite{ALPS20,ALPS_DMFT}

\bibliography{refs_shortened}

\begin{thebibliography}{74}%
\makeatletter
\providecommand \@ifxundefined [1]{%
 \@ifx{#1\undefined}
}%
\providecommand \@ifnum [1]{%
 \ifnum #1\expandafter \@firstoftwo
 \else \expandafter \@secondoftwo
 \fi
}%
\providecommand \@ifx [1]{%
 \ifx #1\expandafter \@firstoftwo
 \else \expandafter \@secondoftwo
 \fi
}%
\providecommand \natexlab [1]{#1}%
\providecommand \enquote  [1]{``#1''}%
\providecommand \bibnamefont  [1]{#1}%
\providecommand \bibfnamefont [1]{#1}%
\providecommand \citenamefont [1]{#1}%
\providecommand \href@noop [0]{\@secondoftwo}%
\providecommand \href [0]{\begingroup \@sanitize@url \@href}%
\providecommand \@href[1]{\@@startlink{#1}\@@href}%
\providecommand \@@href[1]{\endgroup#1\@@endlink}%
\providecommand \@sanitize@url [0]{\catcode `\\12\catcode `\$12\catcode
  `\&12\catcode `\#12\catcode `\^12\catcode `\_12\catcode `\%12\relax}%
\providecommand \@@startlink[1]{}%
\providecommand \@@endlink[0]{}%
\providecommand \url  [0]{\begingroup\@sanitize@url \@url }%
\providecommand \@url [1]{\endgroup\@href {#1}{\urlprefix }}%
\providecommand \urlprefix  [0]{URL }%
\providecommand \Eprint [0]{\href }%
\providecommand \doibase [0]{http://dx.doi.org/}%
\providecommand \selectlanguage [0]{\@gobble}%
\providecommand \bibinfo  [0]{\@secondoftwo}%
\providecommand \bibfield  [0]{\@secondoftwo}%
\providecommand \translation [1]{[#1]}%
\providecommand \BibitemOpen [0]{}%
\providecommand \bibitemStop [0]{}%
\providecommand \bibitemNoStop [0]{.\EOS\space}%
\providecommand \EOS [0]{\spacefactor3000\relax}%
\providecommand \BibitemShut  [1]{\csname bibitem#1\endcsname}%
\let\auto@bib@innerbib\@empty
%</preamble>
\bibitem [{\citenamefont {Bednorz}\ and\ \citenamefont
  {Muller}(1986)}]{Bednorz86}%
  \BibitemOpen
  \bibfield  {author} {\bibinfo {author} {\bibfnamefont {J.}~\bibnamefont
  {Bednorz}}\ and\ \bibinfo {author} {\bibfnamefont {K.}~\bibnamefont
  {Muller}},\ }\href@noop {} {\bibfield  {journal} {\bibinfo  {journal} {Z.
  Phys. B}\ }\textbf {\bibinfo {volume} {64}},\ \bibinfo {pages} {189}
  (\bibinfo {year} {1986})}\BibitemShut {NoStop}%
\bibitem [{\citenamefont {Monthoux}\ \emph {et~al.}(1992)\citenamefont
  {Monthoux}, \citenamefont {Balatsky},\ and\ \citenamefont
  {Pines}}]{Monthoux92}%
  \BibitemOpen
  \bibfield  {author} {\bibinfo {author} {\bibfnamefont {P.}~\bibnamefont
  {Monthoux}}, \bibinfo {author} {\bibfnamefont {A.~V.}\ \bibnamefont
  {Balatsky}}, \ and\ \bibinfo {author} {\bibfnamefont {D.}~\bibnamefont
  {Pines}},\ }\href {\doibase 10.1103/PhysRevB.46.14803} {\bibfield  {journal}
  {\bibinfo  {journal} {Phys. Rev. B}\ }\textbf {\bibinfo {volume} {46}},\
  \bibinfo {pages} {14803} (\bibinfo {year} {1992})}\BibitemShut {NoStop}%
\bibitem [{\citenamefont {Anderson}(1987)}]{Anderson87}%
  \BibitemOpen
  \bibfield  {author} {\bibinfo {author} {\bibfnamefont {P.~W.}\ \bibnamefont
  {Anderson}},\ }\href {\doibase 10.1126/science.235.4793.1196} {\bibfield
  {journal} {\bibinfo  {journal} {Science}\ }\textbf {\bibinfo {volume}
  {235}},\ \bibinfo {pages} {1196} (\bibinfo {year} {1987})}\BibitemShut
  {NoStop}%
\bibitem [{\citenamefont {Chester}(1956)}]{Chester56}%
  \BibitemOpen
  \bibfield  {author} {\bibinfo {author} {\bibfnamefont {G.~V.}\ \bibnamefont
  {Chester}},\ }\href {\doibase 10.1103/PhysRev.103.1693} {\bibfield  {journal}
  {\bibinfo  {journal} {Phys. Rev.}\ }\textbf {\bibinfo {volume} {103}},\
  \bibinfo {pages} {1693} (\bibinfo {year} {1956})}\BibitemShut {NoStop}%
\bibitem [{\citenamefont {Scalapino}(2007)}]{Scalapino07}%
  \BibitemOpen
  \bibfield  {author} {\bibinfo {author} {\bibfnamefont {D.}~\bibnamefont
  {Scalapino}},\ }in\ \href@noop {} {\emph {\bibinfo {booktitle} {Handbook of
  High-Temperature Superconductivity}}},\ \bibinfo {editor} {edited by\
  \bibinfo {editor} {\bibfnamefont {J.}~\bibnamefont {Schrieffer}}\ and\
  \bibinfo {editor} {\bibfnamefont {J.}~\bibnamefont {Brooks}}}\ (\bibinfo
  {publisher} {Springer New York},\ \bibinfo {year} {2007})\ pp.\ \bibinfo
  {pages} {495--526}\BibitemShut {NoStop}%
\bibitem [{\citenamefont {Lee}\ \emph {et~al.}(2006)\citenamefont {Lee},
  \citenamefont {Nagaosa},\ and\ \citenamefont {Wen}}]{Lee07}%
  \BibitemOpen
  \bibfield  {author} {\bibinfo {author} {\bibfnamefont {P.~A.}\ \bibnamefont
  {Lee}}, \bibinfo {author} {\bibfnamefont {N.}~\bibnamefont {Nagaosa}}, \ and\
  \bibinfo {author} {\bibfnamefont {X.-G.}\ \bibnamefont {Wen}},\ }\href@noop
  {} {\bibfield  {journal} {\bibinfo  {journal} {Rev. Mod. Phys.}\ }\textbf
  {\bibinfo {volume} {78}},\ \bibinfo {eid} {17} (\bibinfo {year}
  {2006})}\BibitemShut {NoStop}%
\bibitem [{\citenamefont {Zanchi}\ and\ \citenamefont
  {Schulz}(1996)}]{Zanchi96}%
  \BibitemOpen
  \bibfield  {author} {\bibinfo {author} {\bibfnamefont {D.}~\bibnamefont
  {Zanchi}}\ and\ \bibinfo {author} {\bibfnamefont {H.~J.}\ \bibnamefont
  {Schulz}},\ }\href {\doibase 10.1103/PhysRevB.54.9509} {\bibfield  {journal}
  {\bibinfo  {journal} {Phys. Rev. B}\ }\textbf {\bibinfo {volume} {54}},\
  \bibinfo {pages} {9509} (\bibinfo {year} {1996})}\BibitemShut {NoStop}%
\bibitem [{\citenamefont {Raghu}\ \emph {et~al.}(2010)\citenamefont {Raghu},
  \citenamefont {Kivelson},\ and\ \citenamefont {Scalapino}}]{Raghu10}%
  \BibitemOpen
  \bibfield  {author} {\bibinfo {author} {\bibfnamefont {S.}~\bibnamefont
  {Raghu}}, \bibinfo {author} {\bibfnamefont {S.~A.}\ \bibnamefont {Kivelson}},
  \ and\ \bibinfo {author} {\bibfnamefont {D.~J.}\ \bibnamefont {Scalapino}},\
  }\href {\doibase 10.1103/PhysRevB.81.224505} {\bibfield  {journal} {\bibinfo
  {journal} {Phys. Rev. B}\ }\textbf {\bibinfo {volume} {81}},\ \bibinfo
  {pages} {224505} (\bibinfo {year} {2010})}\BibitemShut {NoStop}%
\bibitem [{\citenamefont {Maier}\ \emph
  {et~al.}(2005{\natexlab{a}})\citenamefont {Maier}, \citenamefont {Jarrell},
  \citenamefont {Schulthess}, \citenamefont {Kent},\ and\ \citenamefont
  {White}}]{Maier05_dwave}%
  \BibitemOpen
  \bibfield  {author} {\bibinfo {author} {\bibfnamefont {T.~A.}\ \bibnamefont
  {Maier}}, \bibinfo {author} {\bibfnamefont {M.}~\bibnamefont {Jarrell}},
  \bibinfo {author} {\bibfnamefont {T.~C.}\ \bibnamefont {Schulthess}},
  \bibinfo {author} {\bibfnamefont {P.~R.~C.}\ \bibnamefont {Kent}}, \ and\
  \bibinfo {author} {\bibfnamefont {J.~B.}\ \bibnamefont {White}},\ }\href
  {\doibase 10.1103/PhysRevLett.95.237001} {\bibfield  {journal} {\bibinfo
  {journal} {Phys. Rev. Lett.}\ }\textbf {\bibinfo {volume} {95}},\ \bibinfo
  {pages} {237001} (\bibinfo {year} {2005}{\natexlab{a}})}\BibitemShut
  {NoStop}%
\bibitem [{\citenamefont {Gull}\ \emph {et~al.}(2009)\citenamefont {Gull},
  \citenamefont {Parcollet}, \citenamefont {Werner},\ and\ \citenamefont
  {Millis}}]{Gull09}%
  \BibitemOpen
  \bibfield  {author} {\bibinfo {author} {\bibfnamefont {E.}~\bibnamefont
  {Gull}}, \bibinfo {author} {\bibfnamefont {O.}~\bibnamefont {Parcollet}},
  \bibinfo {author} {\bibfnamefont {P.}~\bibnamefont {Werner}}, \ and\ \bibinfo
  {author} {\bibfnamefont {A.~J.}\ \bibnamefont {Millis}},\ }\href {\doibase
  10.1103/PhysRevB.80.245102} {\bibfield  {journal} {\bibinfo  {journal} {Phys.
  Rev. B}\ }\textbf {\bibinfo {volume} {80}},\ \bibinfo {pages} {245102}
  (\bibinfo {year} {2009})}\BibitemShut {NoStop}%
\bibitem [{\citenamefont {Gros}(1988)}]{Gros88}%
  \BibitemOpen
  \bibfield  {author} {\bibinfo {author} {\bibfnamefont {C.}~\bibnamefont
  {Gros}},\ }\href {\doibase 10.1103/PhysRevB.38.931} {\bibfield  {journal}
  {\bibinfo  {journal} {Phys. Rev. B}\ }\textbf {\bibinfo {volume} {38}},\
  \bibinfo {pages} {931} (\bibinfo {year} {1988})}\BibitemShut {NoStop}%
\bibitem [{\citenamefont {YOKOYAMA}\ and\ \citenamefont
  {SHIBA}(1988)}]{Yokoyama88}%
  \BibitemOpen
  \bibfield  {author} {\bibinfo {author} {\bibfnamefont {H.}~\bibnamefont
  {YOKOYAMA}}\ and\ \bibinfo {author} {\bibfnamefont {H.}~\bibnamefont
  {SHIBA}},\ }\href {\doibase 10.1143/JPSJ.57.2482} {\bibfield  {journal}
  {\bibinfo  {journal} {JPSJ}\ }\textbf {\bibinfo {volume} {57}},\ \bibinfo
  {pages} {2482} (\bibinfo {year} {1988})}\BibitemShut {NoStop}%
\bibitem [{\citenamefont {Becca}\ \emph {et~al.}(2000)\citenamefont {Becca},
  \citenamefont {Capone},\ and\ \citenamefont {Sorella}}]{Becca00}%
  \BibitemOpen
  \bibfield  {author} {\bibinfo {author} {\bibfnamefont {F.}~\bibnamefont
  {Becca}}, \bibinfo {author} {\bibfnamefont {M.}~\bibnamefont {Capone}}, \
  and\ \bibinfo {author} {\bibfnamefont {S.}~\bibnamefont {Sorella}},\ }\href
  {\doibase 10.1103/PhysRevB.62.12700} {\bibfield  {journal} {\bibinfo
  {journal} {Phys. Rev. B}\ }\textbf {\bibinfo {volume} {62}},\ \bibinfo
  {pages} {12700} (\bibinfo {year} {2000})}\BibitemShut {NoStop}%
\bibitem [{\citenamefont {Paramekanti}\ \emph {et~al.}(2001)\citenamefont
  {Paramekanti}, \citenamefont {Randeria},\ and\ \citenamefont
  {Trivedi}}]{Paramekanti01}%
  \BibitemOpen
  \bibfield  {author} {\bibinfo {author} {\bibfnamefont {A.}~\bibnamefont
  {Paramekanti}}, \bibinfo {author} {\bibfnamefont {M.}~\bibnamefont
  {Randeria}}, \ and\ \bibinfo {author} {\bibfnamefont {N.}~\bibnamefont
  {Trivedi}},\ }\href {\doibase 10.1103/PhysRevLett.87.217002} {\bibfield
  {journal} {\bibinfo  {journal} {Phys. Rev. Lett.}\ }\textbf {\bibinfo
  {volume} {87}},\ \bibinfo {pages} {217002} (\bibinfo {year}
  {2001})}\BibitemShut {NoStop}%
\bibitem [{\citenamefont {Paramekanti}\ \emph {et~al.}(2004)\citenamefont
  {Paramekanti}, \citenamefont {Randeria},\ and\ \citenamefont
  {Trivedi}}]{Paramekanti04}%
  \BibitemOpen
  \bibfield  {author} {\bibinfo {author} {\bibfnamefont {A.}~\bibnamefont
  {Paramekanti}}, \bibinfo {author} {\bibfnamefont {M.}~\bibnamefont
  {Randeria}}, \ and\ \bibinfo {author} {\bibfnamefont {N.}~\bibnamefont
  {Trivedi}},\ }\href {\doibase 10.1103/PhysRevB.70.054504} {\bibfield
  {journal} {\bibinfo  {journal} {Phys. Rev. B}\ }\textbf {\bibinfo {volume}
  {70}},\ \bibinfo {pages} {054504} (\bibinfo {year} {2004})}\BibitemShut
  {NoStop}%
\bibitem [{\citenamefont {Yokoyama}\ \emph {et~al.}(2004)\citenamefont
  {Yokoyama}, \citenamefont {Tanaka}, \citenamefont {Ogata},\ and\
  \citenamefont {Tsuchiura}}]{Yokoyama04}%
  \BibitemOpen
  \bibfield  {author} {\bibinfo {author} {\bibfnamefont {H.}~\bibnamefont
  {Yokoyama}}, \bibinfo {author} {\bibfnamefont {Y.}~\bibnamefont {Tanaka}},
  \bibinfo {author} {\bibfnamefont {M.}~\bibnamefont {Ogata}}, \ and\ \bibinfo
  {author} {\bibfnamefont {H.}~\bibnamefont {Tsuchiura}},\ }\href {\doibase
  10.1143/JPSJ.1119} {\bibfield  {journal} {\bibinfo  {journal} {JPSJ}\
  }\textbf {\bibinfo {volume} {73}},\ \bibinfo {pages} {1119} (\bibinfo {year}
  {2004})}\BibitemShut {NoStop}%
\bibitem [{\citenamefont {{Yokoyama}}\ \emph
  {et~al.}(2012{\natexlab{a}})\citenamefont {{Yokoyama}}, \citenamefont
  {{Ogata}}, \citenamefont {{Tanaka}}, \citenamefont {{Kobayashi}},\ and\
  \citenamefont {{Tsuchiura}}}]{Ogata12a}%
  \BibitemOpen
  \bibfield  {author} {\bibinfo {author} {\bibfnamefont {H.}~\bibnamefont
  {{Yokoyama}}}, \bibinfo {author} {\bibfnamefont {M.}~\bibnamefont {{Ogata}}},
  \bibinfo {author} {\bibfnamefont {Y.}~\bibnamefont {{Tanaka}}}, \bibinfo
  {author} {\bibfnamefont {K.}~\bibnamefont {{Kobayashi}}}, \ and\ \bibinfo
  {author} {\bibfnamefont {H.}~\bibnamefont {{Tsuchiura}}},\ }\href@noop {}
  {\bibfield  {journal} {\bibinfo  {journal} {ArXiv e-prints}\ } (\bibinfo
  {year} {2012}{\natexlab{a}})},\ \Eprint {http://arxiv.org/abs/1208.1102}
  {arXiv:1208.1102 [cond-mat.supr-con]} \BibitemShut {NoStop}%
\bibitem [{\citenamefont {{Yokoyama}}\ \emph
  {et~al.}(2012{\natexlab{b}})\citenamefont {{Yokoyama}}, \citenamefont
  {{Tamura}}, \citenamefont {{Kobayashi}},\ and\ \citenamefont
  {{Ogata}}}]{Ogata12b}%
  \BibitemOpen
  \bibfield  {author} {\bibinfo {author} {\bibfnamefont {H.}~\bibnamefont
  {{Yokoyama}}}, \bibinfo {author} {\bibfnamefont {S.}~\bibnamefont
  {{Tamura}}}, \bibinfo {author} {\bibfnamefont {K.}~\bibnamefont
  {{Kobayashi}}}, \ and\ \bibinfo {author} {\bibfnamefont {M.}~\bibnamefont
  {{Ogata}}},\ }\href@noop {} {\bibfield  {journal} {\bibinfo  {journal} {ArXiv
  e-prints}\ } (\bibinfo {year} {2012}{\natexlab{b}})},\ \Eprint
  {http://arxiv.org/abs/1211.6175} {arXiv:1211.6175 [cond-mat.str-el]}
  \BibitemShut {NoStop}%
\bibitem [{\citenamefont {Haslinger}\ and\ \citenamefont
  {Chubukov}(2003{\natexlab{a}})}]{Haslinger03a}%
  \BibitemOpen
  \bibfield  {author} {\bibinfo {author} {\bibfnamefont {R.}~\bibnamefont
  {Haslinger}}\ and\ \bibinfo {author} {\bibfnamefont {A.~V.}\ \bibnamefont
  {Chubukov}},\ }\href {\doibase 10.1103/PhysRevB.67.140504} {\bibfield
  {journal} {\bibinfo  {journal} {Phys. Rev. B}\ }\textbf {\bibinfo {volume}
  {67}},\ \bibinfo {pages} {140504} (\bibinfo {year}
  {2003}{\natexlab{a}})}\BibitemShut {NoStop}%
\bibitem [{\citenamefont {Haslinger}\ and\ \citenamefont
  {Chubukov}(2003{\natexlab{b}})}]{Haslinger03b}%
  \BibitemOpen
  \bibfield  {author} {\bibinfo {author} {\bibfnamefont {R.}~\bibnamefont
  {Haslinger}}\ and\ \bibinfo {author} {\bibfnamefont {A.~V.}\ \bibnamefont
  {Chubukov}},\ }\href {\doibase 10.1103/PhysRevB.68.214508} {\bibfield
  {journal} {\bibinfo  {journal} {Phys. Rev. B}\ }\textbf {\bibinfo {volume}
  {68}},\ \bibinfo {pages} {214508} (\bibinfo {year}
  {2003}{\natexlab{b}})}\BibitemShut {NoStop}%
\bibitem [{\citenamefont {Yanase}\ and\ \citenamefont
  {Ogata}(2005)}]{Yanase05}%
  \BibitemOpen
  \bibfield  {author} {\bibinfo {author} {\bibfnamefont {Y.}~\bibnamefont
  {Yanase}}\ and\ \bibinfo {author} {\bibfnamefont {M.}~\bibnamefont {Ogata}},\
  }\href {\doibase 10.1143/JPSJ.74.1534} {\bibfield  {journal} {\bibinfo
  {journal} {JPSJ}\ }\textbf {\bibinfo {volume} {74}},\ \bibinfo {pages} {1534}
  (\bibinfo {year} {2005})}\BibitemShut {NoStop}%
\bibitem [{\citenamefont {Benfatto}\ \emph {et~al.}(2006)\citenamefont
  {Benfatto}, \citenamefont {Carbotte},\ and\ \citenamefont
  {Marsiglio}}]{Benfatto06}%
  \BibitemOpen
  \bibfield  {author} {\bibinfo {author} {\bibfnamefont {L.}~\bibnamefont
  {Benfatto}}, \bibinfo {author} {\bibfnamefont {J.~P.}\ \bibnamefont
  {Carbotte}}, \ and\ \bibinfo {author} {\bibfnamefont {F.}~\bibnamefont
  {Marsiglio}},\ }\href {\doibase 10.1103/PhysRevB.74.155115} {\bibfield
  {journal} {\bibinfo  {journal} {Phys. Rev. B}\ }\textbf {\bibinfo {volume}
  {74}},\ \bibinfo {pages} {155115} (\bibinfo {year} {2006})}\BibitemShut
  {NoStop}%
\bibitem [{\citenamefont {Norman}\ \emph {et~al.}(2007)\citenamefont {Norman},
  \citenamefont {Chubukov}, \citenamefont {van Heumen}, \citenamefont
  {Kuzmenko},\ and\ \citenamefont {van~der Marel}}]{Norman07}%
  \BibitemOpen
  \bibfield  {author} {\bibinfo {author} {\bibfnamefont {M.~R.}\ \bibnamefont
  {Norman}}, \bibinfo {author} {\bibfnamefont {A.~V.}\ \bibnamefont
  {Chubukov}}, \bibinfo {author} {\bibfnamefont {E.}~\bibnamefont {van
  Heumen}}, \bibinfo {author} {\bibfnamefont {A.~B.}\ \bibnamefont {Kuzmenko}},
  \ and\ \bibinfo {author} {\bibfnamefont {D.}~\bibnamefont {van~der Marel}},\
  }\href {\doibase 10.1103/PhysRevB.76.220509} {\bibfield  {journal} {\bibinfo
  {journal} {Phys. Rev. B}\ }\textbf {\bibinfo {volume} {76}},\ \bibinfo
  {pages} {220509} (\bibinfo {year} {2007})}\BibitemShut {NoStop}%
\bibitem [{\citenamefont {Marsiglio}\ \emph {et~al.}(2008)\citenamefont
  {Marsiglio}, \citenamefont {van Heumen},\ and\ \citenamefont
  {Kuzmenko}}]{Marsiglio08}%
  \BibitemOpen
  \bibfield  {author} {\bibinfo {author} {\bibfnamefont {F.}~\bibnamefont
  {Marsiglio}}, \bibinfo {author} {\bibfnamefont {E.}~\bibnamefont {van
  Heumen}}, \ and\ \bibinfo {author} {\bibfnamefont {A.~B.}\ \bibnamefont
  {Kuzmenko}},\ }\href {\doibase 10.1103/PhysRevB.77.144510} {\bibfield
  {journal} {\bibinfo  {journal} {Phys. Rev. B}\ }\textbf {\bibinfo {volume}
  {77}},\ \bibinfo {pages} {144510} (\bibinfo {year} {2008})}\BibitemShut
  {NoStop}%
\bibitem [{\citenamefont {Maiti}\ and\ \citenamefont
  {Chubukov}(2010)}]{Maiti10}%
  \BibitemOpen
  \bibfield  {author} {\bibinfo {author} {\bibfnamefont {S.}~\bibnamefont
  {Maiti}}\ and\ \bibinfo {author} {\bibfnamefont {A.~V.}\ \bibnamefont
  {Chubukov}},\ }\href {\doibase 10.1103/PhysRevB.81.245111} {\bibfield
  {journal} {\bibinfo  {journal} {Phys. Rev. B}\ }\textbf {\bibinfo {volume}
  {81}},\ \bibinfo {pages} {245111} (\bibinfo {year} {2010})}\BibitemShut
  {NoStop}%
\bibitem [{\citenamefont {Maier}\ \emph
  {et~al.}(2005{\natexlab{b}})\citenamefont {Maier}, \citenamefont {Jarrell},
  \citenamefont {Pruschke},\ and\ \citenamefont {Hettler}}]{Maier05}%
  \BibitemOpen
  \bibfield  {author} {\bibinfo {author} {\bibfnamefont {T.}~\bibnamefont
  {Maier}}, \bibinfo {author} {\bibfnamefont {M.}~\bibnamefont {Jarrell}},
  \bibinfo {author} {\bibfnamefont {T.}~\bibnamefont {Pruschke}}, \ and\
  \bibinfo {author} {\bibfnamefont {M.~H.}\ \bibnamefont {Hettler}},\ }\href
  {\doibase 10.1103/RevModPhys.77.1027} {\bibfield  {journal} {\bibinfo
  {journal} {Rev. Mod. Phys.}\ }\textbf {\bibinfo {volume} {77}},\ \bibinfo
  {eid} {1027} (\bibinfo {year} {2005}{\natexlab{b}})}\BibitemShut {NoStop}%
\bibitem [{\citenamefont {Maier}\ \emph {et~al.}(2000)\citenamefont {Maier},
  \citenamefont {Jarrell}, \citenamefont {Pruschke},\ and\ \citenamefont
  {Keller}}]{Maier00}%
  \BibitemOpen
  \bibfield  {author} {\bibinfo {author} {\bibfnamefont {T.}~\bibnamefont
  {Maier}}, \bibinfo {author} {\bibfnamefont {M.}~\bibnamefont {Jarrell}},
  \bibinfo {author} {\bibfnamefont {T.}~\bibnamefont {Pruschke}}, \ and\
  \bibinfo {author} {\bibfnamefont {J.}~\bibnamefont {Keller}},\ }\href
  {\doibase 10.1103/PhysRevLett.85.1524} {\bibfield  {journal} {\bibinfo
  {journal} {Phys. Rev. Lett.}\ }\textbf {\bibinfo {volume} {85}},\ \bibinfo
  {pages} {1524} (\bibinfo {year} {2000})}\BibitemShut {NoStop}%
\bibitem [{\citenamefont {Lichtenstein}\ and\ \citenamefont
  {Katsnelson}(2000)}]{Lichtenstein00}%
  \BibitemOpen
  \bibfield  {author} {\bibinfo {author} {\bibfnamefont {A.~I.}\ \bibnamefont
  {Lichtenstein}}\ and\ \bibinfo {author} {\bibfnamefont {M.~I.}\ \bibnamefont
  {Katsnelson}},\ }\href {\doibase 10.1103/PhysRevB.62.R9283} {\bibfield
  {journal} {\bibinfo  {journal} {Phys. Rev. B}\ }\textbf {\bibinfo {volume}
  {62}},\ \bibinfo {pages} {R9283} (\bibinfo {year} {2000})}\BibitemShut
  {NoStop}%
\bibitem [{\citenamefont {Maier}\ \emph {et~al.}(2006)\citenamefont {Maier},
  \citenamefont {Jarrell},\ and\ \citenamefont {Scalapino}}]{Maier06}%
  \BibitemOpen
  \bibfield  {author} {\bibinfo {author} {\bibfnamefont {T.~A.}\ \bibnamefont
  {Maier}}, \bibinfo {author} {\bibfnamefont {M.~S.}\ \bibnamefont {Jarrell}},
  \ and\ \bibinfo {author} {\bibfnamefont {D.~J.}\ \bibnamefont {Scalapino}},\
  }\href {\doibase 10.1103/PhysRevLett.96.047005} {\bibfield  {journal}
  {\bibinfo  {journal} {Phys. Rev. Lett.}\ }\textbf {\bibinfo {volume} {96}},\
  \bibinfo {pages} {047005} (\bibinfo {year} {2006})}\BibitemShut {NoStop}%
\bibitem [{\citenamefont {Maier}\ \emph
  {et~al.}(2007{\natexlab{a}})\citenamefont {Maier}, \citenamefont {Jarrell},\
  and\ \citenamefont {Scalapino}}]{Maier07}%
  \BibitemOpen
  \bibfield  {author} {\bibinfo {author} {\bibfnamefont {T.~A.}\ \bibnamefont
  {Maier}}, \bibinfo {author} {\bibfnamefont {M.}~\bibnamefont {Jarrell}}, \
  and\ \bibinfo {author} {\bibfnamefont {D.~J.}\ \bibnamefont {Scalapino}},\
  }\href {\doibase 10.1103/PhysRevB.75.134519} {\bibfield  {journal} {\bibinfo
  {journal} {Phys. Rev. B}\ }\textbf {\bibinfo {volume} {75}},\ \bibinfo {eid}
  {134519} (\bibinfo {year} {2007}{\natexlab{a}})}\BibitemShut {NoStop}%
\bibitem [{\citenamefont {Maier}\ \emph
  {et~al.}(2007{\natexlab{b}})\citenamefont {Maier}, \citenamefont {Macridin},
  \citenamefont {Jarrell},\ and\ \citenamefont {Scalapino}}]{Maier07B}%
  \BibitemOpen
  \bibfield  {author} {\bibinfo {author} {\bibfnamefont {T.~A.}\ \bibnamefont
  {Maier}}, \bibinfo {author} {\bibfnamefont {A.}~\bibnamefont {Macridin}},
  \bibinfo {author} {\bibfnamefont {M.}~\bibnamefont {Jarrell}}, \ and\
  \bibinfo {author} {\bibfnamefont {D.~J.}\ \bibnamefont {Scalapino}},\ }\href
  {\doibase 10.1103/PhysRevB.76.144516} {\bibfield  {journal} {\bibinfo
  {journal} {Phys. Rev. B}\ }\textbf {\bibinfo {volume} {76}},\ \bibinfo {eid}
  {144516} (\bibinfo {year} {2007}{\natexlab{b}})}\BibitemShut {NoStop}%
\bibitem [{\citenamefont {Civelli}\ \emph {et~al.}(2008)\citenamefont
  {Civelli}, \citenamefont {Capone}, \citenamefont {Georges}, \citenamefont
  {Haule}, \citenamefont {Parcollet}, \citenamefont {Stanescu},\ and\
  \citenamefont {Kotliar}}]{Civelli08}%
  \BibitemOpen
  \bibfield  {author} {\bibinfo {author} {\bibfnamefont {M.}~\bibnamefont
  {Civelli}}, \bibinfo {author} {\bibfnamefont {M.}~\bibnamefont {Capone}},
  \bibinfo {author} {\bibfnamefont {A.}~\bibnamefont {Georges}}, \bibinfo
  {author} {\bibfnamefont {K.}~\bibnamefont {Haule}}, \bibinfo {author}
  {\bibfnamefont {O.}~\bibnamefont {Parcollet}}, \bibinfo {author}
  {\bibfnamefont {T.~D.}\ \bibnamefont {Stanescu}}, \ and\ \bibinfo {author}
  {\bibfnamefont {G.}~\bibnamefont {Kotliar}},\ }\href {\doibase
  10.1103/PhysRevLett.100.046402} {\bibfield  {journal} {\bibinfo  {journal}
  {Phys. Rev. Lett.}\ }\textbf {\bibinfo {volume} {100}},\ \bibinfo {pages}
  {046402} (\bibinfo {year} {2008})}\BibitemShut {NoStop}%
\bibitem [{\citenamefont {Kancharla}\ \emph {et~al.}(2008)\citenamefont
  {Kancharla}, \citenamefont {Kyung}, \citenamefont {S\'en\'echal},
  \citenamefont {Civelli}, \citenamefont {Capone}, \citenamefont {Kotliar},\
  and\ \citenamefont {Tremblay}}]{Kancharla08}%
  \BibitemOpen
  \bibfield  {author} {\bibinfo {author} {\bibfnamefont {S.~S.}\ \bibnamefont
  {Kancharla}}, \bibinfo {author} {\bibfnamefont {B.}~\bibnamefont {Kyung}},
  \bibinfo {author} {\bibfnamefont {D.}~\bibnamefont {S\'en\'echal}}, \bibinfo
  {author} {\bibfnamefont {M.}~\bibnamefont {Civelli}}, \bibinfo {author}
  {\bibfnamefont {M.}~\bibnamefont {Capone}}, \bibinfo {author} {\bibfnamefont
  {G.}~\bibnamefont {Kotliar}}, \ and\ \bibinfo {author} {\bibfnamefont
  {A.-M.~S.}\ \bibnamefont {Tremblay}},\ }\href {\doibase
  10.1103/PhysRevB.77.184516} {\bibfield  {journal} {\bibinfo  {journal} {Phys.
  Rev. B}\ }\textbf {\bibinfo {volume} {77}},\ \bibinfo {pages} {184516}
  (\bibinfo {year} {2008})}\BibitemShut {NoStop}%
\bibitem [{\citenamefont {Maier}\ \emph {et~al.}(2008)\citenamefont {Maier},
  \citenamefont {Poilblanc},\ and\ \citenamefont {Scalapino}}]{Maier08}%
  \BibitemOpen
  \bibfield  {author} {\bibinfo {author} {\bibfnamefont {T.~A.}\ \bibnamefont
  {Maier}}, \bibinfo {author} {\bibfnamefont {D.}~\bibnamefont {Poilblanc}}, \
  and\ \bibinfo {author} {\bibfnamefont {D.~J.}\ \bibnamefont {Scalapino}},\
  }\href {\doibase 10.1103/PhysRevLett.100.237001} {\bibfield  {journal}
  {\bibinfo  {journal} {Phys. Rev. Lett.}\ }\textbf {\bibinfo {volume} {100}},\
  \bibinfo {pages} {237001} (\bibinfo {year} {2008})}\BibitemShut {NoStop}%
\bibitem [{\citenamefont {Civelli}(2009{\natexlab{a}})}]{Civelli09}%
  \BibitemOpen
  \bibfield  {author} {\bibinfo {author} {\bibfnamefont {M.}~\bibnamefont
  {Civelli}},\ }\href {\doibase 10.1103/PhysRevB.79.195113} {\bibfield
  {journal} {\bibinfo  {journal} {Phys. Rev. B}\ }\textbf {\bibinfo {volume}
  {79}},\ \bibinfo {pages} {195113} (\bibinfo {year}
  {2009}{\natexlab{a}})}\BibitemShut {NoStop}%
\bibitem [{\citenamefont {Civelli}(2009{\natexlab{b}})}]{Civelli09b}%
  \BibitemOpen
  \bibfield  {author} {\bibinfo {author} {\bibfnamefont {M.}~\bibnamefont
  {Civelli}},\ }\href {\doibase 10.1103/PhysRevLett.103.136402} {\bibfield
  {journal} {\bibinfo  {journal} {Phys. Rev. Lett.}\ }\textbf {\bibinfo
  {volume} {103}},\ \bibinfo {pages} {136402} (\bibinfo {year}
  {2009}{\natexlab{b}})}\BibitemShut {NoStop}%
\bibitem [{\citenamefont {Sordi}\ \emph {et~al.}(2012)\citenamefont {Sordi},
  \citenamefont {S\'emon}, \citenamefont {Haule},\ and\ \citenamefont
  {Tremblay}}]{Sordi12}%
  \BibitemOpen
  \bibfield  {author} {\bibinfo {author} {\bibfnamefont {G.}~\bibnamefont
  {Sordi}}, \bibinfo {author} {\bibfnamefont {P.}~\bibnamefont {S\'emon}},
  \bibinfo {author} {\bibfnamefont {K.}~\bibnamefont {Haule}}, \ and\ \bibinfo
  {author} {\bibfnamefont {A.-M.~S.}\ \bibnamefont {Tremblay}},\ }\href
  {\doibase 10.1103/PhysRevLett.108.216401} {\bibfield  {journal} {\bibinfo
  {journal} {Phys. Rev. Lett.}\ }\textbf {\bibinfo {volume} {108}},\ \bibinfo
  {pages} {216401} (\bibinfo {year} {2012})}\BibitemShut {NoStop}%
\bibitem [{\citenamefont {{Gull}}\ \emph {et~al.}(2012)\citenamefont {{Gull}},
  \citenamefont {{Parcollet}},\ and\ \citenamefont {{Millis}}}]{Gull12}%
  \BibitemOpen
  \bibfield  {author} {\bibinfo {author} {\bibfnamefont {E.}~\bibnamefont
  {{Gull}}}, \bibinfo {author} {\bibfnamefont {O.}~\bibnamefont {{Parcollet}}},
  \ and\ \bibinfo {author} {\bibfnamefont {A.~J.}\ \bibnamefont {{Millis}}},\
  }\href@noop {} {\enquote {\bibinfo {title} {{Superconductivity and the
  Pseudogap in the two-dimensional Hubbard model}},}\ } (\bibinfo {year}
  {2012}),\ \Eprint {http://arxiv.org/abs/1207.2490} {arXiv:1207.2490
  [cond-mat.str-el]} \BibitemShut {NoStop}%
\bibitem [{\citenamefont {Kozik}\ \emph {et~al.}(2010)\citenamefont {Kozik},
  \citenamefont {Houcke}, \citenamefont {Gull}, \citenamefont {Pollet},
  \citenamefont {Prokof'ev}, \citenamefont {Svistunov},\ and\ \citenamefont
  {Troyer}}]{Kozik10}%
  \BibitemOpen
  \bibfield  {author} {\bibinfo {author} {\bibfnamefont {E.}~\bibnamefont
  {Kozik}}, \bibinfo {author} {\bibfnamefont {K.~V.}\ \bibnamefont {Houcke}},
  \bibinfo {author} {\bibfnamefont {E.}~\bibnamefont {Gull}}, \bibinfo {author}
  {\bibfnamefont {L.}~\bibnamefont {Pollet}}, \bibinfo {author} {\bibfnamefont
  {N.}~\bibnamefont {Prokof'ev}}, \bibinfo {author} {\bibfnamefont
  {B.}~\bibnamefont {Svistunov}}, \ and\ \bibinfo {author} {\bibfnamefont
  {M.}~\bibnamefont {Troyer}},\ }\href {\doibase 10.1209/0295-5075/90/10004}
  {\bibfield  {journal} {\bibinfo  {journal} {Europhys. Lett.}\ }\textbf
  {\bibinfo {volume} {90}},\ \bibinfo {pages} {10004} (\bibinfo {year}
  {2010})}\BibitemShut {NoStop}%
\bibitem [{\citenamefont {Gull}\ \emph {et~al.}(2010)\citenamefont {Gull},
  \citenamefont {Ferrero}, \citenamefont {Parcollet}, \citenamefont {Georges},\
  and\ \citenamefont {Millis}}]{Gull10_clustercompare}%
  \BibitemOpen
  \bibfield  {author} {\bibinfo {author} {\bibfnamefont {E.}~\bibnamefont
  {Gull}}, \bibinfo {author} {\bibfnamefont {M.}~\bibnamefont {Ferrero}},
  \bibinfo {author} {\bibfnamefont {O.}~\bibnamefont {Parcollet}}, \bibinfo
  {author} {\bibfnamefont {A.}~\bibnamefont {Georges}}, \ and\ \bibinfo
  {author} {\bibfnamefont {A.~J.}\ \bibnamefont {Millis}},\ }\href {\doibase
  10.1103/PhysRevB.82.155101} {\bibfield  {journal} {\bibinfo  {journal} {Phys.
  Rev. B}\ }\textbf {\bibinfo {volume} {82}},\ \bibinfo {pages} {155101}
  (\bibinfo {year} {2010})}\BibitemShut {NoStop}%
\bibitem [{\citenamefont {Fuchs}\ \emph {et~al.}(2011)\citenamefont {Fuchs},
  \citenamefont {Gull}, \citenamefont {Pollet}, \citenamefont {Burovski},
  \citenamefont {Kozik}, \citenamefont {Pruschke},\ and\ \citenamefont
  {Troyer}}]{Fuchs11}%
  \BibitemOpen
  \bibfield  {author} {\bibinfo {author} {\bibfnamefont {S.}~\bibnamefont
  {Fuchs}}, \bibinfo {author} {\bibfnamefont {E.}~\bibnamefont {Gull}},
  \bibinfo {author} {\bibfnamefont {L.}~\bibnamefont {Pollet}}, \bibinfo
  {author} {\bibfnamefont {E.}~\bibnamefont {Burovski}}, \bibinfo {author}
  {\bibfnamefont {E.}~\bibnamefont {Kozik}}, \bibinfo {author} {\bibfnamefont
  {T.}~\bibnamefont {Pruschke}}, \ and\ \bibinfo {author} {\bibfnamefont
  {M.}~\bibnamefont {Troyer}},\ }\href@noop {} {\bibfield  {journal} {\bibinfo
  {journal} {Phys. Rev. Lett.}\ }\textbf {\bibinfo {volume} {106}},\ \bibinfo
  {pages} {030401} (\bibinfo {year} {2011})}\BibitemShut {NoStop}%
\bibitem [{\citenamefont {Sakai}\ \emph {et~al.}(2012)\citenamefont {Sakai},
  \citenamefont {Sangiovanni}, \citenamefont {Civelli}, \citenamefont {Motome},
  \citenamefont {Held},\ and\ \citenamefont {Imada}}]{Sakai12}%
  \BibitemOpen
  \bibfield  {author} {\bibinfo {author} {\bibfnamefont {S.}~\bibnamefont
  {Sakai}}, \bibinfo {author} {\bibfnamefont {G.}~\bibnamefont {Sangiovanni}},
  \bibinfo {author} {\bibfnamefont {M.}~\bibnamefont {Civelli}}, \bibinfo
  {author} {\bibfnamefont {Y.}~\bibnamefont {Motome}}, \bibinfo {author}
  {\bibfnamefont {K.}~\bibnamefont {Held}}, \ and\ \bibinfo {author}
  {\bibfnamefont {M.}~\bibnamefont {Imada}},\ }\href {\doibase
  10.1103/PhysRevB.85.035102} {\bibfield  {journal} {\bibinfo  {journal} {Phys.
  Rev. B}\ }\textbf {\bibinfo {volume} {85}},\ \bibinfo {pages} {035102}
  (\bibinfo {year} {2012})}\BibitemShut {NoStop}%
\bibitem [{\citenamefont {Gull}\ \emph {et~al.}(2008)\citenamefont {Gull},
  \citenamefont {Werner}, \citenamefont {Parcollet},\ and\ \citenamefont
  {Troyer}}]{Gull08}%
  \BibitemOpen
  \bibfield  {author} {\bibinfo {author} {\bibfnamefont {E.}~\bibnamefont
  {Gull}}, \bibinfo {author} {\bibfnamefont {P.}~\bibnamefont {Werner}},
  \bibinfo {author} {\bibfnamefont {O.}~\bibnamefont {Parcollet}}, \ and\
  \bibinfo {author} {\bibfnamefont {M.}~\bibnamefont {Troyer}},\ }\href
  {\doibase 10.1209/0295-5075/82/57003} {\bibfield  {journal} {\bibinfo
  {journal} {Europhys. Lett.}\ }\textbf {\bibinfo {volume} {82}},\ \bibinfo
  {pages} {57003} (\bibinfo {year} {2008})}\BibitemShut {NoStop}%
\bibitem [{\citenamefont {Gull}\ \emph
  {et~al.}(2011{\natexlab{a}})\citenamefont {Gull}, \citenamefont {Millis},
  \citenamefont {Lichtenstein}, \citenamefont {Rubtsov}, \citenamefont
  {Troyer},\ and\ \citenamefont {Werner}}]{Gull11}%
  \BibitemOpen
  \bibfield  {author} {\bibinfo {author} {\bibfnamefont {E.}~\bibnamefont
  {Gull}}, \bibinfo {author} {\bibfnamefont {A.~J.}\ \bibnamefont {Millis}},
  \bibinfo {author} {\bibfnamefont {A.~I.}\ \bibnamefont {Lichtenstein}},
  \bibinfo {author} {\bibfnamefont {A.~N.}\ \bibnamefont {Rubtsov}}, \bibinfo
  {author} {\bibfnamefont {M.}~\bibnamefont {Troyer}}, \ and\ \bibinfo {author}
  {\bibfnamefont {P.}~\bibnamefont {Werner}},\ }\href {\doibase
  10.1103/RevModPhys.83.349} {\bibfield  {journal} {\bibinfo  {journal} {Rev.
  Mod. Phys.}\ }\textbf {\bibinfo {volume} {83}},\ \bibinfo {pages} {349}
  (\bibinfo {year} {2011}{\natexlab{a}})}\BibitemShut {NoStop}%
\bibitem [{\citenamefont {Gull}\ \emph
  {et~al.}(2011{\natexlab{b}})\citenamefont {Gull}, \citenamefont {Staar},
  \citenamefont {Fuchs}, \citenamefont {Nukala}, \citenamefont {Summers},
  \citenamefont {Pruschke}, \citenamefont {Schulthess},\ and\ \citenamefont
  {Maier}}]{Gull10_submatrix}%
  \BibitemOpen
  \bibfield  {author} {\bibinfo {author} {\bibfnamefont {E.}~\bibnamefont
  {Gull}}, \bibinfo {author} {\bibfnamefont {P.}~\bibnamefont {Staar}},
  \bibinfo {author} {\bibfnamefont {S.}~\bibnamefont {Fuchs}}, \bibinfo
  {author} {\bibfnamefont {P.}~\bibnamefont {Nukala}}, \bibinfo {author}
  {\bibfnamefont {M.~S.}\ \bibnamefont {Summers}}, \bibinfo {author}
  {\bibfnamefont {T.}~\bibnamefont {Pruschke}}, \bibinfo {author}
  {\bibfnamefont {T.~C.}\ \bibnamefont {Schulthess}}, \ and\ \bibinfo {author}
  {\bibfnamefont {T.}~\bibnamefont {Maier}},\ }\href {\doibase
  10.1103/PhysRevB.83.075122} {\bibfield  {journal} {\bibinfo  {journal} {Phys.
  Rev. B}\ }\textbf {\bibinfo {volume} {83}},\ \bibinfo {pages} {075122}
  (\bibinfo {year} {2011}{\natexlab{b}})}\BibitemShut {NoStop}%
\bibitem [{\citenamefont {Huscroft}\ \emph {et~al.}(2001)\citenamefont
  {Huscroft}, \citenamefont {Jarrell}, \citenamefont {Maier}, \citenamefont
  {Moukouri},\ and\ \citenamefont {Tahvildarzadeh}}]{Huscroft01}%
  \BibitemOpen
  \bibfield  {author} {\bibinfo {author} {\bibfnamefont {C.}~\bibnamefont
  {Huscroft}}, \bibinfo {author} {\bibfnamefont {M.}~\bibnamefont {Jarrell}},
  \bibinfo {author} {\bibfnamefont {T.}~\bibnamefont {Maier}}, \bibinfo
  {author} {\bibfnamefont {S.}~\bibnamefont {Moukouri}}, \ and\ \bibinfo
  {author} {\bibfnamefont {A.~N.}\ \bibnamefont {Tahvildarzadeh}},\ }\href
  {\doibase 10.1103/PhysRevLett.86.139} {\bibfield  {journal} {\bibinfo
  {journal} {Phys. Rev. Lett.}\ }\textbf {\bibinfo {volume} {86}},\ \bibinfo
  {pages} {139} (\bibinfo {year} {2001})}\BibitemShut {NoStop}%
\bibitem [{\citenamefont {Parcollet}\ \emph {et~al.}(2004)\citenamefont
  {Parcollet}, \citenamefont {Biroli},\ and\ \citenamefont
  {Kotliar}}]{Parcollet04}%
  \BibitemOpen
  \bibfield  {author} {\bibinfo {author} {\bibfnamefont {O.}~\bibnamefont
  {Parcollet}}, \bibinfo {author} {\bibfnamefont {G.}~\bibnamefont {Biroli}}, \
  and\ \bibinfo {author} {\bibfnamefont {G.}~\bibnamefont {Kotliar}},\ }\href
  {\doibase 10.1103/PhysRevLett.92.226402} {\bibfield  {journal} {\bibinfo
  {journal} {Phys. Rev. Lett.}\ }\textbf {\bibinfo {volume} {92}},\ \bibinfo
  {pages} {226402} (\bibinfo {year} {2004})}\BibitemShut {NoStop}%
\bibitem [{\citenamefont {Macridin}\ \emph {et~al.}(2006)\citenamefont
  {Macridin}, \citenamefont {Jarrell}, \citenamefont {Maier}, \citenamefont
  {Kent},\ and\ \citenamefont {D'Azevedo}}]{Macridin06}%
  \BibitemOpen
  \bibfield  {author} {\bibinfo {author} {\bibfnamefont {A.}~\bibnamefont
  {Macridin}}, \bibinfo {author} {\bibfnamefont {M.}~\bibnamefont {Jarrell}},
  \bibinfo {author} {\bibfnamefont {T.}~\bibnamefont {Maier}}, \bibinfo
  {author} {\bibfnamefont {P.~R.~C.}\ \bibnamefont {Kent}}, \ and\ \bibinfo
  {author} {\bibfnamefont {E.}~\bibnamefont {D'Azevedo}},\ }\href {\doibase
  10.1103/PhysRevLett.97.036401} {\bibfield  {journal} {\bibinfo  {journal}
  {Phys. Rev. Lett.}\ }\textbf {\bibinfo {volume} {97}},\ \bibinfo {eid}
  {036401} (\bibinfo {year} {2006})}\BibitemShut {NoStop}%
\bibitem [{\citenamefont {Werner}\ \emph {et~al.}(2009)\citenamefont {Werner},
  \citenamefont {Gull}, \citenamefont {Parcollet},\ and\ \citenamefont
  {Millis}}]{Werner098site}%
  \BibitemOpen
  \bibfield  {author} {\bibinfo {author} {\bibfnamefont {P.}~\bibnamefont
  {Werner}}, \bibinfo {author} {\bibfnamefont {E.}~\bibnamefont {Gull}},
  \bibinfo {author} {\bibfnamefont {O.}~\bibnamefont {Parcollet}}, \ and\
  \bibinfo {author} {\bibfnamefont {A.~J.}\ \bibnamefont {Millis}},\ }\href
  {\doibase 10.1103/PhysRevB.80.045120} {\bibfield  {journal} {\bibinfo
  {journal} {Phys. Rev. B}\ }\textbf {\bibinfo {volume} {80}},\ \bibinfo {eid}
  {045120} (\bibinfo {year} {2009})}\BibitemShut {NoStop}%
\bibitem [{\citenamefont {Sakai}\ \emph {et~al.}(2009)\citenamefont {Sakai},
  \citenamefont {Motome},\ and\ \citenamefont {Imada}}]{Sakai09}%
  \BibitemOpen
  \bibfield  {author} {\bibinfo {author} {\bibfnamefont {S.}~\bibnamefont
  {Sakai}}, \bibinfo {author} {\bibfnamefont {Y.}~\bibnamefont {Motome}}, \
  and\ \bibinfo {author} {\bibfnamefont {M.}~\bibnamefont {Imada}},\ }\href
  {\doibase 10.1103/PhysRevLett.102.056404} {\bibfield  {journal} {\bibinfo
  {journal} {Phys. Rev. Lett.}\ }\textbf {\bibinfo {volume} {102}},\ \bibinfo
  {pages} {056404} (\bibinfo {year} {2009})}\BibitemShut {NoStop}%
\bibitem [{\citenamefont {Sakai}\ \emph {et~al.}(2010)\citenamefont {Sakai},
  \citenamefont {Motome},\ and\ \citenamefont {Imada}}]{Sakai10}%
  \BibitemOpen
  \bibfield  {author} {\bibinfo {author} {\bibfnamefont {S.}~\bibnamefont
  {Sakai}}, \bibinfo {author} {\bibfnamefont {Y.}~\bibnamefont {Motome}}, \
  and\ \bibinfo {author} {\bibfnamefont {M.}~\bibnamefont {Imada}},\ }\href
  {\doibase 10.1103/PhysRevB.82.134505} {\bibfield  {journal} {\bibinfo
  {journal} {Phys. Rev. B}\ }\textbf {\bibinfo {volume} {82}},\ \bibinfo
  {pages} {134505} (\bibinfo {year} {2010})}\BibitemShut {NoStop}%
\bibitem [{\citenamefont {Sordi}\ \emph {et~al.}(2010)\citenamefont {Sordi},
  \citenamefont {Haule},\ and\ \citenamefont {Tremblay}}]{Sordi10}%
  \BibitemOpen
  \bibfield  {author} {\bibinfo {author} {\bibfnamefont {G.}~\bibnamefont
  {Sordi}}, \bibinfo {author} {\bibfnamefont {K.}~\bibnamefont {Haule}}, \ and\
  \bibinfo {author} {\bibfnamefont {A.-M.~S.}\ \bibnamefont {Tremblay}},\
  }\href {\doibase 10.1103/PhysRevLett.104.226402} {\bibfield  {journal}
  {\bibinfo  {journal} {Phys. Rev. Lett.}\ }\textbf {\bibinfo {volume} {104}},\
  \bibinfo {pages} {226402} (\bibinfo {year} {2010})}\BibitemShut {NoStop}%
\bibitem [{\citenamefont {Yang}\ \emph {et~al.}(2011)\citenamefont {Yang},
  \citenamefont {Fotso}, \citenamefont {Su}, \citenamefont {Galanakis},
  \citenamefont {Khatami}, \citenamefont {She}, \citenamefont {Moreno},
  \citenamefont {Zaanen},\ and\ \citenamefont {Jarrell}}]{Yang11}%
  \BibitemOpen
  \bibfield  {author} {\bibinfo {author} {\bibfnamefont {S.-X.}\ \bibnamefont
  {Yang}}, \bibinfo {author} {\bibfnamefont {H.}~\bibnamefont {Fotso}},
  \bibinfo {author} {\bibfnamefont {S.-Q.}\ \bibnamefont {Su}}, \bibinfo
  {author} {\bibfnamefont {D.}~\bibnamefont {Galanakis}}, \bibinfo {author}
  {\bibfnamefont {E.}~\bibnamefont {Khatami}}, \bibinfo {author} {\bibfnamefont
  {J.-H.}\ \bibnamefont {She}}, \bibinfo {author} {\bibfnamefont
  {J.}~\bibnamefont {Moreno}}, \bibinfo {author} {\bibfnamefont
  {J.}~\bibnamefont {Zaanen}}, \ and\ \bibinfo {author} {\bibfnamefont
  {M.}~\bibnamefont {Jarrell}},\ }\href {\doibase
  10.1103/PhysRevLett.106.047004} {\bibfield  {journal} {\bibinfo  {journal}
  {Phys. Rev. Lett.}\ }\textbf {\bibinfo {volume} {106}},\ \bibinfo {pages}
  {047004} (\bibinfo {year} {2011})}\BibitemShut {NoStop}%
\bibitem [{\citenamefont {Sordi}\ \emph {et~al.}(2011)\citenamefont {Sordi},
  \citenamefont {Haule},\ and\ \citenamefont {Tremblay}}]{Sordi11}%
  \BibitemOpen
  \bibfield  {author} {\bibinfo {author} {\bibfnamefont {G.}~\bibnamefont
  {Sordi}}, \bibinfo {author} {\bibfnamefont {K.}~\bibnamefont {Haule}}, \ and\
  \bibinfo {author} {\bibfnamefont {A.-M.~S.}\ \bibnamefont {Tremblay}},\
  }\href {\doibase 10.1103/PhysRevB.84.075161} {\bibfield  {journal} {\bibinfo
  {journal} {Phys. Rev. B}\ }\textbf {\bibinfo {volume} {84}},\ \bibinfo
  {pages} {075161} (\bibinfo {year} {2011})}\BibitemShut {NoStop}%
\bibitem [{\citenamefont {Maier}\ \emph {et~al.}(2004)\citenamefont {Maier},
  \citenamefont {Jarrell}, \citenamefont {Macridin},\ and\ \citenamefont
  {Slezak}}]{Maier04}%
  \BibitemOpen
  \bibfield  {author} {\bibinfo {author} {\bibfnamefont {T.~A.}\ \bibnamefont
  {Maier}}, \bibinfo {author} {\bibfnamefont {M.}~\bibnamefont {Jarrell}},
  \bibinfo {author} {\bibfnamefont {A.}~\bibnamefont {Macridin}}, \ and\
  \bibinfo {author} {\bibfnamefont {C.}~\bibnamefont {Slezak}},\ }\href
  {\doibase 10.1103/PhysRevLett.92.027005} {\bibfield  {journal} {\bibinfo
  {journal} {Phys. Rev. Lett.}\ }\textbf {\bibinfo {volume} {92}},\ \bibinfo
  {pages} {027005} (\bibinfo {year} {2004})}\BibitemShut {NoStop}%
\bibitem [{\citenamefont {Haule}\ and\ \citenamefont
  {Kotliar}(2007)}]{Haule07}%
  \BibitemOpen
  \bibfield  {author} {\bibinfo {author} {\bibfnamefont {K.}~\bibnamefont
  {Haule}}\ and\ \bibinfo {author} {\bibfnamefont {G.}~\bibnamefont
  {Kotliar}},\ }\href {\doibase 10.1209/0295-5075/77/27007} {\bibfield
  {journal} {\bibinfo  {journal} {Europhysics Letters}\ }\textbf {\bibinfo
  {volume} {77}},\ \bibinfo {pages} {27007} (\bibinfo {year}
  {2007})}\BibitemShut {NoStop}%
\bibitem [{\citenamefont {Singh}(2007)}]{Singh07}%
  \BibitemOpen
  \bibfield  {author} {\bibinfo {author} {\bibfnamefont {D.~J.}\ \bibnamefont
  {Singh}},\ }\href {\doibase 10.1103/PhysRevB.75.012501} {\bibfield  {journal}
  {\bibinfo  {journal} {Phys. Rev. B}\ }\textbf {\bibinfo {volume} {75}},\
  \bibinfo {pages} {012501} (\bibinfo {year} {2007})}\BibitemShut {NoStop}%
\bibitem [{\citenamefont {Affleck}\ \emph {et~al.}(1988)\citenamefont
  {Affleck}, \citenamefont {Zou}, \citenamefont {Hsu},\ and\ \citenamefont
  {Anderson}}]{Affleck88}%
  \BibitemOpen
  \bibfield  {author} {\bibinfo {author} {\bibfnamefont {I.}~\bibnamefont
  {Affleck}}, \bibinfo {author} {\bibfnamefont {Z.}~\bibnamefont {Zou}},
  \bibinfo {author} {\bibfnamefont {T.}~\bibnamefont {Hsu}}, \ and\ \bibinfo
  {author} {\bibfnamefont {P.~W.}\ \bibnamefont {Anderson}},\ }\href {\doibase
  10.1103/PhysRevB.38.745} {\bibfield  {journal} {\bibinfo  {journal} {Phys.
  Rev. B}\ }\textbf {\bibinfo {volume} {38}},\ \bibinfo {pages} {745} (\bibinfo
  {year} {1988})}\BibitemShut {NoStop}%
\bibitem [{\citenamefont {Norman}\ \emph {et~al.}(2000)\citenamefont {Norman},
  \citenamefont {Randeria}, \citenamefont {Janko},\ and\ \citenamefont
  {Campuzano}}]{Norman00}%
  \BibitemOpen
  \bibfield  {author} {\bibinfo {author} {\bibfnamefont {M.~R.}\ \bibnamefont
  {Norman}}, \bibinfo {author} {\bibfnamefont {M.}~\bibnamefont {Randeria}},
  \bibinfo {author} {\bibfnamefont {B.}~\bibnamefont {Janko}}, \ and\ \bibinfo
  {author} {\bibfnamefont {J.~C.}\ \bibnamefont {Campuzano}},\ }\href@noop {}
  {\bibfield  {journal} {\bibinfo  {journal} {Phys. Rev. B}\ }\textbf {\bibinfo
  {volume} {61}},\ \bibinfo {pages} {14742} (\bibinfo {year}
  {2000})}\BibitemShut {NoStop}%
\bibitem [{\citenamefont {van~der Marel}\ \emph {et~al.}(2002)\citenamefont
  {van~der Marel}, \citenamefont {Leggett}, \citenamefont {Loram},\ and\
  \citenamefont {Kirtley}}]{Marel02}%
  \BibitemOpen
  \bibfield  {author} {\bibinfo {author} {\bibfnamefont {D.}~\bibnamefont
  {van~der Marel}}, \bibinfo {author} {\bibfnamefont {A.~J.}\ \bibnamefont
  {Leggett}}, \bibinfo {author} {\bibfnamefont {J.~W.}\ \bibnamefont {Loram}},
  \ and\ \bibinfo {author} {\bibfnamefont {J.~R.}\ \bibnamefont {Kirtley}},\
  }\href {\doibase 10.1103/PhysRevB.66.140501} {\bibfield  {journal} {\bibinfo
  {journal} {Phys. Rev. B}\ }\textbf {\bibinfo {volume} {66}},\ \bibinfo
  {pages} {140501} (\bibinfo {year} {2002})}\BibitemShut {NoStop}%
\bibitem [{\citenamefont {Maldague}(1977)}]{Maldague77}%
  \BibitemOpen
  \bibfield  {author} {\bibinfo {author} {\bibfnamefont {P.~F.}\ \bibnamefont
  {Maldague}},\ }\href {\doibase 10.1103/PhysRevB.16.2437} {\bibfield
  {journal} {\bibinfo  {journal} {Phys. Rev. B}\ }\textbf {\bibinfo {volume}
  {16}},\ \bibinfo {pages} {2437/2446} (\bibinfo {year} {1977})}\BibitemShut
  {NoStop}%
\bibitem [{\citenamefont {Baeriswyl}\ \emph {et~al.}(1987)\citenamefont
  {Baeriswyl}, \citenamefont {Gros},\ and\ \citenamefont {Rice}}]{Baeriswyl87}%
  \BibitemOpen
  \bibfield  {author} {\bibinfo {author} {\bibfnamefont {D.}~\bibnamefont
  {Baeriswyl}}, \bibinfo {author} {\bibfnamefont {C.}~\bibnamefont {Gros}}, \
  and\ \bibinfo {author} {\bibfnamefont {T.~M.}\ \bibnamefont {Rice}},\ }\href
  {\doibase 10.1103/PhysRevB.35.8391} {\bibfield  {journal} {\bibinfo
  {journal} {Phys. Rev. B}\ }\textbf {\bibinfo {volume} {35}},\ \bibinfo
  {pages} {8391/8395} (\bibinfo {year} {1987})}\BibitemShut {NoStop}%
\bibitem [{\citenamefont {Millis}\ and\ \citenamefont
  {Coppersmith}(1990)}]{Millis90}%
  \BibitemOpen
  \bibfield  {author} {\bibinfo {author} {\bibfnamefont {A.~J.}\ \bibnamefont
  {Millis}}\ and\ \bibinfo {author} {\bibfnamefont {S.~N.}\ \bibnamefont
  {Coppersmith}},\ }\href {\doibase 10.1103/PhysRevB.42.10807} {\bibfield
  {journal} {\bibinfo  {journal} {Phys. Rev. B}\ }\textbf {\bibinfo {volume}
  {42}},\ \bibinfo {pages} {10807/10810} (\bibinfo {year} {1990})}\BibitemShut
  {NoStop}%
\bibitem [{\citenamefont {Hirsch}(1993)}]{Hirsch93}%
  \BibitemOpen
  \bibfield  {author} {\bibinfo {author} {\bibfnamefont {J.}~\bibnamefont
  {Hirsch}},\ }\href {\doibase 10.1016/0022-3697(93)90150-P} {\bibfield
  {journal} {\bibinfo  {journal} {Journal of the Physics and Chemistry of
  Solids}\ }\textbf {\bibinfo {volume} {54}},\ \bibinfo {pages} {1101}
  (\bibinfo {year} {1993})},\ \bibinfo {note} {conference on Spectroscopies in
  Novel Superconductors, Santa Fe, NM, Mar 17-19, 1993}\BibitemShut {NoStop}%
\bibitem [{\citenamefont {Hirsch}\ and\ \citenamefont
  {Marsiglio}(2000)}]{Hirsch00}%
  \BibitemOpen
  \bibfield  {author} {\bibinfo {author} {\bibfnamefont {J.}~\bibnamefont
  {Hirsch}}\ and\ \bibinfo {author} {\bibfnamefont {F.}~\bibnamefont
  {Marsiglio}},\ }\href {\doibase 10.1016/S0921-4534(99)00669-3} {\bibfield
  {journal} {\bibinfo  {journal} {Physica C: Superconductivity}\ }\textbf
  {\bibinfo {volume} {331}},\ \bibinfo {pages} {150 } (\bibinfo {year}
  {2000})}\BibitemShut {NoStop}%
\bibitem [{\citenamefont {Chakravarty}\ \emph {et~al.}(1999)\citenamefont
  {Chakravarty}, \citenamefont {Kee},\ and\ \citenamefont
  {Abrahams}}]{Chakravarty98}%
  \BibitemOpen
  \bibfield  {author} {\bibinfo {author} {\bibfnamefont {S.}~\bibnamefont
  {Chakravarty}}, \bibinfo {author} {\bibfnamefont {H.-Y.}\ \bibnamefont
  {Kee}}, \ and\ \bibinfo {author} {\bibfnamefont {E.}~\bibnamefont
  {Abrahams}},\ }\href {\doibase 10.1103/PhysRevLett.82.2366} {\bibfield
  {journal} {\bibinfo  {journal} {Phys. Rev. Lett.}\ }\textbf {\bibinfo
  {volume} {82}},\ \bibinfo {pages} {2366} (\bibinfo {year}
  {1999})}\BibitemShut {NoStop}%
\bibitem [{\citenamefont {Anderson}(1998)}]{Anderson98}%
  \BibitemOpen
  \bibfield  {author} {\bibinfo {author} {\bibfnamefont {P.~W.}\ \bibnamefont
  {Anderson}},\ }\href {\doibase 10.1126/science.279.5354.1196} {\bibfield
  {journal} {\bibinfo  {journal} {Science}\ }\textbf {\bibinfo {volume}
  {279}},\ \bibinfo {pages} {1196} (\bibinfo {year} {1998})}\BibitemShut
  {NoStop}%
\bibitem [{\citenamefont {Molegraaf}\ \emph {et~al.}(2002)\citenamefont
  {Molegraaf}, \citenamefont {Presura}, \citenamefont {van~der Marel},
  \citenamefont {Kes},\ and\ \citenamefont {Li}}]{Molegraaf02}%
  \BibitemOpen
  \bibfield  {author} {\bibinfo {author} {\bibfnamefont {H.~J.~A.}\
  \bibnamefont {Molegraaf}}, \bibinfo {author} {\bibfnamefont {C.}~\bibnamefont
  {Presura}}, \bibinfo {author} {\bibfnamefont {D.}~\bibnamefont {van~der
  Marel}}, \bibinfo {author} {\bibfnamefont {P.~H.}\ \bibnamefont {Kes}}, \
  and\ \bibinfo {author} {\bibfnamefont {M.}~\bibnamefont {Li}},\ }\href
  {\doibase 10.1126/science.1069947} {\bibfield  {journal} {\bibinfo  {journal}
  {Science}\ }\textbf {\bibinfo {volume} {295}},\ \bibinfo {pages} {2239}
  (\bibinfo {year} {2002})}\BibitemShut {NoStop}%
\bibitem [{\citenamefont {Santander-Syro}\ \emph {et~al.}(2002)\citenamefont
  {Santander-Syro}, \citenamefont {Lobo}, \citenamefont {Bontemps},
  \citenamefont {Konstantinovic}, \citenamefont {Li},\ and\ \citenamefont
  {Raffy}}]{Santander02}%
  \BibitemOpen
  \bibfield  {author} {\bibinfo {author} {\bibfnamefont {A.~F.}\ \bibnamefont
  {Santander-Syro}}, \bibinfo {author} {\bibfnamefont {R.}~\bibnamefont
  {Lobo}}, \bibinfo {author} {\bibfnamefont {N.}~\bibnamefont {Bontemps}},
  \bibinfo {author} {\bibfnamefont {Z.}~\bibnamefont {Konstantinovic}},
  \bibinfo {author} {\bibfnamefont {Z.}~\bibnamefont {Li}}, \ and\ \bibinfo
  {author} {\bibfnamefont {H.}~\bibnamefont {Raffy}},\ }\href {\doibase
  10.1103/PhysRevLett.88.097005} {\bibfield  {journal} {\bibinfo  {journal}
  {Phys. Rev. Lett.}\ }\textbf {\bibinfo {volume} {88}},\ \bibinfo {pages}
  {097005} (\bibinfo {year} {2002})}\BibitemShut {NoStop}%
\bibitem [{\citenamefont {Santander-Syro}\ \emph {et~al.}(2004)\citenamefont
  {Santander-Syro}, \citenamefont {Lobo}, \citenamefont {Bontemps},
  \citenamefont {Lopera}, \citenamefont {Girat\'a}, \citenamefont
  {Konstantinovic}, \citenamefont {Li},\ and\ \citenamefont
  {Raffy}}]{Santander04}%
  \BibitemOpen
  \bibfield  {author} {\bibinfo {author} {\bibfnamefont {A.~F.}\ \bibnamefont
  {Santander-Syro}}, \bibinfo {author} {\bibfnamefont {R.~P. S.~M.}\
  \bibnamefont {Lobo}}, \bibinfo {author} {\bibfnamefont {N.}~\bibnamefont
  {Bontemps}}, \bibinfo {author} {\bibfnamefont {W.}~\bibnamefont {Lopera}},
  \bibinfo {author} {\bibfnamefont {D.}~\bibnamefont {Girat\'a}}, \bibinfo
  {author} {\bibfnamefont {Z.}~\bibnamefont {Konstantinovic}}, \bibinfo
  {author} {\bibfnamefont {Z.~Z.}\ \bibnamefont {Li}}, \ and\ \bibinfo {author}
  {\bibfnamefont {H.}~\bibnamefont {Raffy}},\ }\href {\doibase
  10.1103/PhysRevB.70.134504} {\bibfield  {journal} {\bibinfo  {journal} {Phys.
  Rev. B}\ }\textbf {\bibinfo {volume} {70}},\ \bibinfo {pages} {134504}
  (\bibinfo {year} {2004})}\BibitemShut {NoStop}%
\bibitem [{\citenamefont {Deutscher}\ \emph {et~al.}(2005)\citenamefont
  {Deutscher}, \citenamefont {Santander-Syro},\ and\ \citenamefont
  {Bontemps}}]{Deutscher05}%
  \BibitemOpen
  \bibfield  {author} {\bibinfo {author} {\bibfnamefont {G.}~\bibnamefont
  {Deutscher}}, \bibinfo {author} {\bibfnamefont {A.~F.}\ \bibnamefont
  {Santander-Syro}}, \ and\ \bibinfo {author} {\bibfnamefont {N.}~\bibnamefont
  {Bontemps}},\ }\href {\doibase 10.1103/PhysRevB.72.092504} {\bibfield
  {journal} {\bibinfo  {journal} {Phys. Rev. B}\ }\textbf {\bibinfo {volume}
  {72}},\ \bibinfo {pages} {092504} (\bibinfo {year} {2005})}\BibitemShut
  {NoStop}%
\bibitem [{\citenamefont {Carbone}\ \emph {et~al.}(2006)\citenamefont
  {Carbone}, \citenamefont {Kuzmenko}, \citenamefont {Molegraaf}, \citenamefont
  {van Heumen}, \citenamefont {Lukovac}, \citenamefont {Marsiglio},
  \citenamefont {van~der Marel}, \citenamefont {Haule}, \citenamefont
  {Kotliar}, \citenamefont {Berger}, \citenamefont {Courjault}, \citenamefont
  {Kes},\ and\ \citenamefont {Li}}]{Carbone06}%
  \BibitemOpen
  \bibfield  {author} {\bibinfo {author} {\bibfnamefont {F.}~\bibnamefont
  {Carbone}}, \bibinfo {author} {\bibfnamefont {A.~B.}\ \bibnamefont
  {Kuzmenko}}, \bibinfo {author} {\bibfnamefont {H.~J.~A.}\ \bibnamefont
  {Molegraaf}}, \bibinfo {author} {\bibfnamefont {E.}~\bibnamefont {van
  Heumen}}, \bibinfo {author} {\bibfnamefont {V.}~\bibnamefont {Lukovac}},
  \bibinfo {author} {\bibfnamefont {F.}~\bibnamefont {Marsiglio}}, \bibinfo
  {author} {\bibfnamefont {D.}~\bibnamefont {van~der Marel}}, \bibinfo {author}
  {\bibfnamefont {K.}~\bibnamefont {Haule}}, \bibinfo {author} {\bibfnamefont
  {G.}~\bibnamefont {Kotliar}}, \bibinfo {author} {\bibfnamefont
  {H.}~\bibnamefont {Berger}}, \bibinfo {author} {\bibfnamefont
  {S.}~\bibnamefont {Courjault}}, \bibinfo {author} {\bibfnamefont {P.~H.}\
  \bibnamefont {Kes}}, \ and\ \bibinfo {author} {\bibfnamefont
  {M.}~\bibnamefont {Li}},\ }\href {\doibase 10.1103/PhysRevB.74.064510}
  {\bibfield  {journal} {\bibinfo  {journal} {Phys. Rev. B}\ }\textbf {\bibinfo
  {volume} {74}},\ \bibinfo {pages} {064510} (\bibinfo {year}
  {2006})}\BibitemShut {NoStop}%
\bibitem [{\citenamefont {Bauer}\ \emph {et~al.}(2011)\citenamefont {Bauer}
  \emph {et~al.}}]{ALPS20}%
  \BibitemOpen
  \bibfield  {author} {\bibinfo {author} {\bibfnamefont {B.}~\bibnamefont
  {Bauer}} \emph {et~al.},\ }\href@noop {} {\bibfield  {journal} {\bibinfo
  {journal} {Journal of Statistical Mechanics: Theory and Experiment}\ }\textbf
  {\bibinfo {volume} {2011}},\ \bibinfo {pages} {P05001} (\bibinfo {year}
  {2011})}\BibitemShut {NoStop}%
\bibitem [{\citenamefont {Gull}\ \emph
  {et~al.}(2011{\natexlab{c}})\citenamefont {Gull}, \citenamefont {Werner},
  \citenamefont {Fuchs}, \citenamefont {Surer}, \citenamefont {Pruschke},\ and\
  \citenamefont {Troyer}}]{ALPS_DMFT}%
  \BibitemOpen
  \bibfield  {author} {\bibinfo {author} {\bibfnamefont {E.}~\bibnamefont
  {Gull}}, \bibinfo {author} {\bibfnamefont {P.}~\bibnamefont {Werner}},
  \bibinfo {author} {\bibfnamefont {S.}~\bibnamefont {Fuchs}}, \bibinfo
  {author} {\bibfnamefont {B.}~\bibnamefont {Surer}}, \bibinfo {author}
  {\bibfnamefont {T.}~\bibnamefont {Pruschke}}, \ and\ \bibinfo {author}
  {\bibfnamefont {M.}~\bibnamefont {Troyer}},\ }\href {\doibase
  10.1016/j.cpc.2010.12.050} {\bibfield  {journal} {\bibinfo  {journal}
  {Computer Physics Communications}\ }\textbf {\bibinfo {volume} {182}},\
  \bibinfo {pages} {1078 } (\bibinfo {year} {2011}{\natexlab{c}})}\BibitemShut
  {NoStop}%
\end{thebibliography}%
\end{document}